\documentclass[11pt]{article}
\usepackage{amsmath,amssymb,bm,epsf,epsfig,graphicx}
\input epsf.sty
\usepackage[utf8]{inputenc}
\usepackage[english]{babel}
\usepackage{amsmath}
\usepackage{latexsym}
\usepackage{amsfonts}
\usepackage{mathtools}
\usepackage{amssymb}
\usepackage[left=3cm,right=3cm,top=3.1cm,bottom=3.1cm]{geometry}
\usepackage{cite}
\usepackage{tensor}
\usepackage{simplewick}
\usepackage{hyperref}
\usepackage{stackengine}
\numberwithin{equation}{section}

\def\bra#1{\langle #1 |}

\def\aver#1{\left\langle\, #1 \,\right\rangle}

\def\p{\partial}
\def \N {\mathcal{ N}}

\def \be {\begin{eqnarray}}
\def \ee {\end{eqnarray}}
\def \bdm {\begin{displaymath}}
\def \edm {\end{displaymath}}

\def \tr{{\rm tr}}

\def\del {\partial}
\def\0{\nonumber}

\def \VV {{\mathbb V}}
\def \WW {{\mathbb W}}

\def \CC {{\mathbb C}}
\def \HH {{\mathbb H}}

\def\RRR{\mathfrak{R}}
\def\VVV{\mathcal{V}}

\newcommand{\e}[0]{\mathrm{e}}

\newcommand{\mc}[1]{{\mathcal{#1}}}

\newcommand{\V}[0]{\mathbb{V}}

\newcommand{\Mean}[1]{\left\langle #1 \right\rangle}

\usepackage{xparse, marginnote}
\usepackage[usenames, dvipsnames, svgnames, table]{xcolor}
\newcounter{draftcommentcnt}
\NewDocumentCommand{\draftcomment}{s O{red} m}{%
	\def\margnote{\IfBooleanTF{#1}{\marginnote}{\marginpar}}%
	\stepcounter{draftcommentcnt}%
	\textcolor{#2}{#3}%
	\margnote{\textcolor{#2}{$\Leftarrow$ \arabic{draftcommentcnt}}}%
}

\begin{document}

\vspace*{3.1cm}

\centerline{\Large \bf Localization of effective actions}\vspace{.4cm}
\centerline{\Large \bf in  Heterotic String  Field Theory} \vspace{.2cm}
% \centerline{\large \bf in open string field theory}
\vspace*{.1cm}

\begin{center}

{\large Harold Erbin$^{(a)}$\footnote{Email: harold.erbin at gmail.com}, Carlo Maccaferri$^{(a)}$\footnote{Email: maccafer at gmail.com} and Jakub Vo\v{s}mera$^{(b)}$\footnote{Email: vosmera at gmail.com} }
\vskip 1 cm
$^{(a)}${\it Dipartimento di Fisica, Universit\`a di Torino, \\INFN  Sezione di Torino and Arnold-Regge Center\\
Via Pietro Giuria 1, I-10125 Torino, Italy}
\vskip .5 cm
$^{(b)}${\it Institute of Physics of the AS CR, Na Slovance 2,\\  and Institute of Particle and Nuclear Physics, Charles University,\\ V Hole\v{s}ovi\v{c}k\'{a}ch 2,  Prague 8, Czech~Republic}

\end{center}

\vspace*{6.0ex}

\centerline{\bf Abstract}
\bigskip
We consider the algebraic couplings in the tree level effective action of the heterotic string. We show how these couplings can be computed from closed string field theory. When the light fields we are interested in are charged under an underlying $\N=2$ $R$-charge in the left-moving sector, their quartic effective potential localizes at the boundary of the worldsheet moduli space, in complete analogy to the previously studied open string case. In particular we are able to  compute the quartic closed string field theory potential without resorting to any explicit expression for  the 3- and the 4-strings vertices  but only using the $L_\infty$ relations between them. As a non trivial example we show how the heterotic Yang-Mills quartic potential arises in this way. \vfill \eject

\baselineskip=16pt

\tableofcontents
\newpage
\section{Introduction and Summary}\label{intro}
String field theory (SFT) is an  approach to  non-perturbative string theory, which has the perturbative world-sheet description in its basic definition.
Many progresses on different fronts have been achieved in the past years. Complete constructions of  covariant superstring actions have become  available \cite{Moosavian:2019ydz, Kunitomo:2019glq, Erler:2017onq, Konopka:2016grr, Erler:2016ybs, Sen:2015uaa,  Kunitomo:2015usa}. On an independent flow, evidence for background independence has emerged, thanks to the plethora of discovered  classical solutions  (mainly in the case of open strings) and related aspects \cite{Erler:2019fye, Vosmera:2019mzw, Maccaferri:2019ogq, Mattiello:2019gxc, Maccaferri:2018vwo, Erler:2019nmz,  Erler:2019xof, Kudrna:2018mxa, Cho:2018nfn, Kojita:2016jwe,  Kudrna:2016ack,  Maccaferri:2015cha,  Erler:2014eqa, Maccaferri:2014cpa,Kudrna:2014rya,Erler:2013wda,  Kudrna:2012re,Erler:2012qn,  Murata:2011ep, Hata:2011ke, Kiermaier:2010cf, Bonora:2010hi, Schnabl:2005gv
}.  In parallel we have also realized, mainly thanks to the work of Ashoke Sen, that a  consistent approach to superstring perturbation theory necessarily needs a SFT formulation  \cite{Sen:2019jpm, deLacroix:2018tml, Sen:2017szq, Sen:2016qap,Sen:2016uzq, Sen:2016gqt, Pius:2016jsl, Sen-restoration, Sen:2015hha, Sen:2014dqa, Sen:2014pia
}. See \cite{Erler:2019vhl} and \cite{deLacroix:2017lif} for recent reviews on the above-mentioned achievements.

 A quantum consistent theory necessarily includes closed strings as dynamical degrees of freedom and in fact,  from this perspective, the simplest model to consider is
  a theory with only closed strings since, at least perturbatively (i.e. in absence of D-branes), open strings cannot be created by closed strings interacting between themselves. However closed string field theory (see\cite{Erler:2019loq } for a recent review) is not as explicit as open string field theory because the fundamental vertices defining its interactions necessarily include integrations over implicitly-defined internal regions of the moduli space of punctured Riemann surfaces, together with  local coordinates around punctures, for which we generally do not have closed form expressions. But since these data are necessary for constructing  off-shell amplitudes, it seems that a direct approach towards analytic computations in closed string field theory is still not really available, although   progress in this direction is happening \cite{Cho:2019anu, Moosavian:2019pmd, Costello:2019fuh, Headrick:2018ncs, Moosavian:2017sev, Moosavian:2017qsp}.
However, on second thought, we would expect that  when we are concerned with physical quantities, the final result should be independent on the various un-physical data that are needed for the definition of the fundamental vertices.  Then it should be possible to compute observables  by-passing the explicit knowledge of the off-shell data, only using the fundamental properties that ensure consistency of the action, such as gauge invariance.

In this paper we  make a step in this direction  and we consider  the tree level effective action of the Heterotic string \cite{Cai:1986sa, Gross:1986mw, Bergshoeff:1989de}. To do so we start from the WZW-like formulation of heterotic string field theory \cite{Okawa:2004ii,Berkovits:2004xh} in the NS sector and we classically integrate out the massive fields by solving their equation of motion in terms of the massless fields, in  analogy to what has been done in the early days by Berkovits and Schnabl \cite{BS} for the open superstring. More precisely, we will be interested in the algebraic couplings of the effective action, which define the field-potential. Then,  following  \cite{Maccaferri:2018vwo, Maccaferri:2019ogq}, we observe that when the massless  fields we are interested in  are charged under the $R$-charge of an underlying $\N=2$ SCFT in the left-moving sector, the full quartic potential localizes at the boundary of moduli space of the 4-punctured sphere.  Although we are here talking about closed string amplitudes, the mechanism is completely parallel  to the open string case analyzed in \cite{Maccaferri:2018vwo} and \cite{Maccaferri:2019ogq, Vosmera:2019mzw}.
%Beside the  $\N=2$ structure the only other needed ingredients are the $L_\infty$ relations which indeed guarantees gauge invariance at the classical level. The basic reason behind this simplification is that conservation of $J$ charge localizes the quartic potential to the boundary of  moduli space of the 4-punctured sphere and all the complicated details of the elementary quartic vertex are instead located very far from the boundary where the quartic potential has zero support.

Summarizing with a bit more of details, our results gives formulae which are tantalizingly almost identical to \cite{Maccaferri:2018vwo}. In the space of dynamical heterotic NS string fields in the large Hilbert space, collectively indicated as $\Phi$, we split the string field in the kernel of $L_0^+=L_0+\overline L_0$ and its complement using the corresponding projectors $P_0$ and $1-P_0$
\be
\Phi=P_0\Phi+(1-P_0)\Phi\equiv \varphi+\RRR.
\ee
%\be
%P_0=\lim_{\epsilon\to0}\frac{\epsilon}{L_0^+ + \epsilon}.
%\ee
%

These projectors are  useful to correctly account for how the Siegel-gauge propagator defines a contracting homotopy operator for the BRST charge
\be
\left[Q,\frac{b_0^+}{L_0^+}\right]=1-P_0\equiv\overline P_0.\label{homotopy}
\ee
Then, the effective action for the massless field $\varphi$ is explicitly given, up to quartic order, as
\begin{equation}
	\label{eq:eff-action-light-exp}
	\begin{aligned}
	S_{\text{eff}}(\varphi)
		&
		= \frac{1}{2} \Mean{\eta_0 \varphi, Q \varphi}
			+ \frac{1}{3!} \Mean{\eta_0 \varphi, [ \varphi, Q \varphi ]}+S_{\text{eff}}^{(4)}(\varphi)+O(\varphi^5),
				\end{aligned}
\end{equation}
where the quartic coupling is given by
\be
S_{\text{eff}}^{(4)}(\varphi)&=& \frac{1}{4!} \Mean{\eta_0 \varphi, \big[\varphi, [\varphi, Q \varphi]\big]}
			+ \frac{1}{4!} \Mean{\eta_0 \varphi, [ \varphi, Q \varphi,Q\varphi]}	\0	\\ &&
- \frac{1}{8} \Mean{ [\eta_0 \varphi, Q \varphi],\xi_0\frac{b_0^+}{L_0^+}   \overline P_0\, [\eta_0 \varphi, Q \varphi ]}\label{Seff4}
\ee
and it contains, in the propagator term, the contribution from the massive fields which have been (classically) integrated out. As anticipated, computing this quantity for a generic field in the kernel of $L_0^+$ requires  off-shell data which are in general not known in a concrete enough form for us to obtain a final analytic result in terms of spacetime fields.
However if we are interested in purely algebraic couplings we can go further. At zero momentum there are two kinds of physical fields: the standard physical fields (which we will denote $\varphi_A$) and the ghost dilaton, $\varphi_D$. The latter is given by $$\varphi_D(z,\overline z)=D\,c\xi \del\xi e^{-2\phi}Q(\del c-\overline\del\overline c)(z,\overline z)=D\,c \gamma^{-2} Q(\del c-\overline\del\overline c)(z,\overline z),\,\quad\quad D\in \mathbb{R}$$ and it does not depend on the matter conformal field theory.
 Then all of the other physical fields  are of the form
% \footnote{In fact the only massless physical field which is not captured by this ansatz is the heterotic ghost dilaton $$\varphi_D(z,\overline z)=D\,c\xi \del\xi e^{-2\phi}Q(\del c-\overline\del\overline c)(z,\overline z)=D\,c \gamma^{-2} Q(\del c-\overline\del\overline c)(z,\overline z),$$ which does not depend on the matter conformal field theory and therefore is not sensible to the localization mechanism we are dealing with in this paper, see section \eqref{ker}.}
\be
\varphi_A(z,\overline z)=\varepsilon_{ik}\,c \xi e^{-\phi}\VV^i_{\frac12}(z)\,\overline c\overline\WW^k_1(\overline z)\equiv c\gamma^{-1}(z)\VVV_{\frac12,1}(z,\overline z)\,\overline c(\overline z)\label{phi-phys},
\ee
where $\overline\WW^k_1$'s  are anti-holomorphic conformal primaries of weight 1 living in the bosonic right-moving  part and $\VV^i_{\frac12}(z)$ are   $\N=1$ superconformal primaries of weight 1/2 in the supersymmetric left-moving sector. The generic polarization $\varepsilon_{ik}$ spans the remaining degrees of freedom at zero momentum.

As it turns out only  $\varphi_A$  can contribute to the elementary cubic couplings in \eqref{eq:eff-action-light-exp} and the ghost dilaton decouples.
On the other hand, for quartic couplings, we don't have an analytic way to approach the ghost dilaton, but we can still say quite a lot on the quartic couplings of  $\varphi_A$ alone, which are the main target of this work.%For what concerns  quartic couplings we will only focus on the computation of the effective couplings of the $\varphi_A$ fields alone, leaving for the future to study the effect of the ghost dilaton at quartic order.

Concentrating thus on the $\varphi_A$ fields only, we further restrict our attention to $\N=1$ superconformal primaries which can be written as sum of short $\N=2$ primaries of $R$-charge $\pm1$ \cite{Sen-restoration, Maccaferri:2018vwo, Maccaferri:2019ogq, Vosmera:2019mzw}
\be
\VV^j_{\frac12}(z)=(\VV^{j}_{\frac12})^+(z)+(\VV^{j}_{\frac12})^-(z)
\ee
and accordingly from \eqref{phi-phys}
\be
\VVV_{\frac12,1}(z, \overline z)&=&\VVV^{+}_{\frac12,1}(z,\overline z)+\VVV^{-}_{\frac12,1}(z,\overline z),\\
\varphi_A&=&\varphi^++\varphi^-.
\ee
As we will see, the first consequence of this decomposition is that cubic couplings will be identically vanishing
\be
S_{\text{eff}}^{(3)}\left(\varphi^++\varphi^-\right)=0.
\ee
Secondly and most importantly,  the conservation of $R$-charge, together with the $L_\infty$ relations \eqref{eq:Linf3}, and the fundamental fact \eqref{homotopy} allow to re-write the quartic potential as
\begin{equation}
	\begin{aligned}
	S_{\text{eff}}^{(4)}(\varphi^++\varphi^-)
		=
			&
			-\frac{1}{8} \Mean{[ \varphi^-, \eta_0\varphi^-],\, P_0[\varphi^+, Q \varphi^+]}
			- \frac{1}{8} \Mean{ [\varphi^-, \varphi^+],\,  P_0[\eta_0 \varphi^-, Q \varphi^+]}			\\ &
			+(+\leftrightarrow -).	\end{aligned}\label{Seffloc}
\end{equation}
Thanks to the presence of the projector $P_0\sim \lim_{t\to\infty}e^{-t L_0^+}$, the full quartic potential is thus captured by a SFT diagram where 2 three-vertices are connected by an infinitely long cylinder (corresponding to the proper-time $t \to \infty$ limit of the closed string propagator $1/L_0^+=\int_0^\infty dt e^{-t L_0^+}$). This is a contribution at the boundary of moduli space of the 4-punctured sphere, where punctures collide two by two.  In particular the 2-string products of primary fields in the kernel of $L_0^+$ are  reduced, thanks to the $P_0$ in front, to simple Fock states which are obtained by leading-order OPE
\be
P_0[\Phi_1,\Phi_2]=b_0^-\delta(L_0^-)\{\Phi_1\,\Phi_2\}_{0,0}(0,0)|0\rangle_{SL(2,\CC)},
\ee
where by $\{A\,B\}_{k,\overline k}$ we denote the field which is found in the symmetric OPE at the order of singularity $z^{-k}\overline z^{-\overline k}$.
Thanks to this $enormous$ simplification the $\varphi_A$-effective action can be evaluated via elementary two-point functions to give a universal expression in the matter sector
\be
S_{\text{eff}}(\varphi^++\varphi^-)=\frac14\Big(\bra{\HH_{1,1}^{+}}\,\HH_{1,1}^{-}\rangle + \bra{\HH_{0,1}}\,\HH_{0,1}\rangle \Big)+\mathcal{O}(\varphi_A^5),\label{seffmatt}
\ee
where we have defined the charged heterotic auxiliary fields as
\be
\HH_{1,1}^{\pm}(z,\overline z)\equiv\lim_{\epsilon\to 0} \,(2\overline\epsilon)\,\VVV^{\pm}_{\frac12,1}(z+\epsilon,\overline z+\overline\epsilon)\, \VVV^{\pm}_{\frac12,1}(z-\epsilon,\overline z-\overline\epsilon),
\ee
with $R$-charge $\pm2$ and the neutral one
\be
\HH_{0,1}(z,\overline z)\equiv\lim_{\epsilon\to 0}\, |2\epsilon|^2 \,\VVV^{+}_{\frac12,1}(z+\epsilon,\overline z+\overline\epsilon)\, \VVV^{-}_{\frac12,1}(z-\epsilon,\overline z-\overline\epsilon),
\ee
with vanishing $R$-charge.
It is important to realize that while in the open string case localization was just a shortcut to a more complicated computation involving (still exactly computable)  4-point OSFT diagrams \cite{BS, Vosmera:2019mzw}, here this is a truly major advantage as it gives us the possibility to access to the full quartic potential without  an explicit representation of the fundamental vertices of the microscopic closed string field theory; all complications of closed string field theory have been by-passed. This is the central result of this paper, which is organized as follows.

In section \ref{sec2} we review the construction  of NS Heterotic string field theory in the large Hilbert space as originally defined in \cite{Okawa:2004ii,Berkovits:2004xh}. We then split the string field into massless and massive components and we perturbatively integrate out the latter. Here, improving from \cite{Maccaferri:2018vwo, Maccaferri:2019ogq}, we carefully analyze the gauge fixing procedure for the massive fields, in particular we show that, once the gauge-fixed equation of motion for the massive fields are solved in terms of the massless fields, then the out-of-gauge equations of motion for the massive fields are  satisfied if   the massless equation of motion are also satisfied. This allows us to state that whenever we are interested in the dynamics of the massless fields, the full string field theory equation of motion is controlled by the massless equations  of motion only. Our analysis is  performed up to cubic order in the massless fields for the equation of motion and up to quartic order for the action. A complete result to all orders will be presented in \cite{Erbin:2019xx}  in the case of generic SFT actions whose interactions are organized into a cyclic $L_\infty$ structure, as it typically happens in small Hilbert space theories. The effective action for the massless fields is then simply obtained by substituting the gauge fixed solution for the massive fields into the original classical action. Then in section \ref{sec3}, after having shown how to reduce the kernel of $L_0^+$ to the physical fields only, we concentrate on the quartic effective potential of fields in the form \eqref{phi-phys}, which admit an $\N=2$ decomposition. We show by manipulations in the large Hilbert space analogous to \cite{Maccaferri:2018vwo}  that the full quartic potential reduces to \eqref{Seffloc}. In section \ref{sec4} we show the consequence of this localization mechanism in flat $D=10$ space-time where we correctly obtain the  Yang-Mills potential of the heterotic gauge field. Moreover we confirm the absence of algebraic couplings for the metric and the Kalb-Ramond field since the correspoding auxiliary fields are indentically vanishing. We end up in section \ref{sec5} with conclusions  and future perspectives. Appendix \ref{app:A} contains the detailed  derivation of how the gauge constraints for the massive fields are controlled by the massless equation of motion, up to quartic order. Appendix \ref{app:B} shows that we can consistently ignore all of the
fields in the kernel of $L_0^+$ which are not physical, either by simply setting them to zero  (for states annihilated  by $\eta_0$) or integrating them out (for states proportional to $c_0^+$), without producing new couplings up to quartic order.
Finally appendix \ref{app:C} contains the proof that there are no cubic couplings for physical fields with  $\N=2$ $R$-charge equal to $\pm1$.

%We can farther decompose the massless zero momentum field $\hat\varphi$ as a physical field $\varphi$ plus the Nakanishi-Lautrup field $\tilde\varphi$
%\be
%\hat\varphi=\varphi+\tilde\varphi.
%\ee
%The physical field can be universally written as
%\be
%\varphi(z,\overline z)=\varepsilon_{ik}\,c \xi e^{-\phi}\VV^i_{\frac12}(z)\,\overline c\overline\WW^k_1(\overline z),
%\ee
%where $\overline\WW^k_1$'s  are anti-holomorphic conformal primaries of weight 1 living in the bosonic right-moving  part and $\VV^i_{\frac12}(z)$ are   $\N=1$ superconformal primaries of weight 1/2 in the supersymmetric left-moving sector, and the generic tensor $\varepsilon_{ik}$ spans the zero momentum degrees of freedom.
%
%The Nakanishi-Lautrup field is the only other zero momentum closed-string state  in the kernel of $L_0^+$  and it is given by
%\be
%\tilde\varphi(z,\overline z)=B  c\del c\xi\del\xi e^{-2\phi}(z) \overline\del\overline c(\overline z),
%\ee
%where $B$ is the corresponding zero-momentum space-time field.

%%%%%%%%%%%%%%%%%%%%%%%%%%%%%%%%%
\section{Heterotic effective action}\label{sec2}
%%%%%%%%%%%%%%%%%%%%%%%%%%%%
In this section we first review the construction of heterotic string field theory in the large Hilbert space and then we will integrate out the massive fields
to get the tree level effective action for the massless fields, up to quartic order.

\subsection{Heterotic String Field Theory}
\label{sec:MargDef}

The action for heterotic SFT in the large Hilbert space  \cite{Okawa:2004ii,Berkovits:2004xh} expanded up to quartic order in the NS sector is
\begin{subequations}
\label{eq:Shet}
\begin{align}
    S(\Phi) &= \frac{1}{2}\big\langle \eta_0 \Phi,Q\Phi\big\rangle+\frac{\kappa}{3!}\big\langle \eta_0 \Phi,[\Phi,Q\Phi]\big\rangle+\nonumber\\
    &\hspace{2cm}+\frac{\kappa^2}{4!}\bigg(\big\langle \eta_0 \Phi,[\Phi,Q\Phi,Q\Phi]\big\rangle+\big\langle\eta_0 \Phi,[\Phi,[\Phi,Q\Phi]]\big\rangle\bigg)+\mathcal{O}(\kappa^3)\\
    &=\frac{1}{2}\big\langle \eta_0 \Phi,Q\Phi\big\rangle+\mathcal{I}(\Phi),
\end{align}
\end{subequations}
with $Q=Q_B+\overline{Q}_B$ where
\begin{subequations}
\begin{align}
Q_B &= \oint \frac{dz}{2\pi i}\bigg\{c (T_\mathrm{m}+T_{\xi\eta}+T_\phi)+c\p c b +\eta e^{\phi}G - \eta \p \eta e^{2\phi}b\bigg\}
\,,\\
\overline{Q}_B &= \oint \frac{dz}{2\pi i}\bigg\{\overline{c} \overline{T}_\mathrm{m}+\overline{c}\overline{\p} \overline{c} \overline{b}\bigg\}\,,
\end{align}
\end{subequations}
where by $G$ we have denoted the matter supercurrent. We have also defined the interacting part
\begin{align}
\mathcal{I}(\Phi)&=\frac{\kappa}{3!}\big\langle \eta_0 \Phi,[\Phi,Q\Phi]\big\rangle+\nonumber\\
    &\hspace{2cm}+\frac{\kappa^2}{4!}\bigg(\big\langle \eta_0 \Phi,[\Phi,Q\Phi,Q\Phi]\big\rangle+\big\langle\eta_0 \Phi,[\Phi,[\Phi,Q\Phi]]\big\rangle\bigg)+\mathcal{O}(\kappa^3)\,.
\end{align}
The string field $\Phi$ is a combination of  states consisting in the left moving sector  of an $\mathcal{N}=1$ matter SCFT with central charge $c=15$ together with the $(b,c)$ and $(\beta,\gamma)$ systems while, in the right-moving sector, of a $\overline{c}=26$ matter CFT together with the $(\overline{b},\overline{c})$ system. The string field $\Phi$ carries ghost number $+1$ and picture number $0$ and satisfies the level matching conditions
\begin{align}
    b_0^-\Phi = L_0^- \Phi =0\,,
\end{align}
where $b_0^- = b_0-\overline{b}_0$, $L_0^- = L_0-\overline{L}_0$. Following  Zwiebach's standard notation \cite{Zwiebach:1992ie},
the inner product is defined as $\langle A,B\rangle = \langle A|c_0^-|B\rangle$ where $c_0^- = \frac{1}{2}(c_0-\overline{c}_0)$. The multi string products $[\Phi_1,\ldots ,\Phi_n]$ are all graded-commutative $wrt$ grassmanality, for instance, $[A,B]=(-1)^{AB}[B,A]$ and similarly for the higher products. In addition we have
\begin{subequations}
\begin{align}
   \big \langle A,B\big\rangle &= (-1)^{(A+1)(B+1)}\big\langle B,A\big\rangle\,,\\
   \big \langle QA,B\big\rangle &= (-1)^{A}\big\langle A,QB\big\rangle\,,\\
   \big \langle \eta_0 A,B\big\rangle &= (-1)^{A}\big\langle A,\eta_0 B\big\rangle\,,\\
   \big \langle [A,B],C\big\rangle &= (-1)^{A+B}\big\langle A,[B,C]\big\rangle\,.
\end{align}
\end{subequations}
%Note that we can also define the multilinear forms
%\begin{align}
%    \big\{\Phi_1,\ldots, \Phi_n \big\} = \big\langle \Phi_1,[\Phi_2,\ldots,\Phi_n]\big\rangle\,,
%\end{align}
%which are all graded-commutative.
In order to guarantee that the action is invariant under an appropriate non-linear gauge transformation \cite{Okawa:2004ii,Berkovits:2004xh}, the multi-string products satisfy the relations of an $L_\infty$ algebra which, up to cubic order are explicitly
\begingroup\allowdisplaybreaks
\begin{subequations}
\begin{align}
    0&= Q^2 \,,\\
    0&=Q[\Phi_1,\Phi_2]+[Q\Phi_1,\Phi_2]+(-1)^{\Phi_1}[\Phi_1,Q\Phi_2]\,,\\
    0&= Q[\Phi_1,\Phi_2,\Phi_3]+[Q\Phi_1,\Phi_2,\Phi_3]+(-1)^{\Phi_1}[\Phi_1,Q\Phi_2,\Phi_3]+(-1)^{\Phi_1+\Phi_2}[\Phi_1,\Phi_2,Q\Phi_3]+\nonumber\\
    &\hspace{4cm}+(-1)^{\Phi_1}[\Phi_1,[\Phi_2,\Phi_3]]+(-1)^{\Phi_2(\Phi_1+1)+\Phi_1\Phi_3}[\Phi_2,[\Phi_3,\Phi_1]]+\nonumber\\
    &\hspace{8cm}+(-1)^{ \Phi_3(\Phi_1+\Phi_2+1)} [\Phi_3,[\Phi_1,\Phi_2]]\,.\label{eq:Linf3}\\
    &\hspace{0.2cm}\vdots\nonumber
\end{align}
\end{subequations}
\endgroup
%Varying the action \eqref{eq:Shet}, we obtain the classical equations of motion
%\begin{subequations}
%\label{eq:EOM4}
%\begin{align}
%    \text{EOM}(\Phi)&= -\eta_0 Q\Phi+\frac{\kappa}{2!} [\eta_0 \Phi,Q\Phi]+\frac{\kappa^2}{3!}\bigg(
%   [\eta_0 \Phi,Q\Phi,Q\Phi]+\nonumber\\
%   &\hspace{2cm}
%   +[\eta_0 \Phi,[\Phi,Q\Phi]]
%   -\frac{1}{2}[\Phi,[\Phi,\eta_0 Q\Phi]]
%   -\frac{1}{2}[ \Phi,[Q\Phi,\eta_0 \Phi]]
%\bigg)+\mathcal{O}(\kappa^3)\\
%&= -\eta_0 Q\Phi+\mathcal{J}(\Phi)\,,
%\end{align}
%\end{subequations}
%where we have defined the interaction term
%\begin{align}
%\mathcal{J}(\Phi) &= +\frac{\kappa}{2!} [\eta_0 \Phi,Q\Phi]+\frac{\kappa^2}{3!}\bigg(
%   [\eta_0 \Phi,Q\Phi,Q\Phi]+\nonumber\\
%   &\hspace{2cm}
%   +[\eta_0 \Phi,[\Phi,Q\Phi]]
%   -\frac{1}{2}[\Phi,[\Phi,\eta_0 Q\Phi]]
%   -\frac{1}{2}[ \Phi,[Q\Phi,\eta_0 \Phi]]
%\bigg)+\mathcal{O}(\kappa^3)\,.
%\end{align}
From now one we will set the gravitational coupling $\kappa$ equal to unity,
\be
\kappa=1.
\ee

\subsection{Effective action}

Let us now focus on deriving the effective action for the massless modes of the heterotic SFT.
%We will therefore specialize to the case when $P=P_0$ is the projector on $\text{ker}\,L_0^+$ where $L_0^+=L_0+\overline{L}_0$.
%Recall that the full action of the heterotic SFT is
%\begin{align}
%S(\Phi)=\frac{1}{2}\big\langle \eta_0 \Phi,Q\Phi\big\rangle+\mathcal{I}(\Phi)\,,\label{eq:Shet2}
%\end{align}
%where the interaction part $\mathcal{I}(\Phi)$ reads
%\begin{align}
%\mathcal{I}(\Phi)&=\frac{1}{3!}\big\langle \eta_0 \Phi,[\Phi,Q\Phi]\big\rangle+\nonumber\\
%    &\hspace{2cm}+\frac{1}{4!}\bigg(\big\langle \eta_0 \Phi,[\Phi,Q\Phi,Q\Phi]\big\rangle+\big\langle\eta_0 \Phi,[\Phi,[\Phi,Q\Phi]]\big\rangle\bigg)+\mathcal{O}(\Phi^5)\,.
%\end{align}
Varying the action \eqref{eq:Shet} with respect to $\Phi$, we obtain the equation of motion
\begin{align}
\text{EOM}(\Phi) = -\eta_0 Q \Phi + \mathcal{J}(\Phi)\,,
\end{align}
where we have separated the interacting part as
\begin{align}
\mathcal{J}(\Phi) &=+\frac{1}{2!} [\eta_0 \Phi,Q\Phi]+\frac{1}{3!}\bigg(
   [\eta_0 \Phi,Q\Phi,Q\Phi]+\nonumber\\
   &\hspace{2cm}
   +[\eta_0 \Phi,[\Phi,Q\Phi]]
   -\frac{1}{2}[\Phi,[\Phi,\eta_0 Q\Phi]]
   -\frac{1}{2}[ \Phi,[Q\Phi,\eta_0 \Phi]]
\bigg)+\mathcal{O}(\Phi^4)\,.
\end{align}
To isolate the massless modes (in view of our latest interest in the zero momentum sector) we consider the projector into the kernel of $L_0^+$, $P_0$ (together with $\overline P_0\equiv1-P_0$) and we
 split the string field as
%\begin{subequations}
\begin{align}
\Phi &=P_0\Phi+\overline P_0\Phi\equiv \varphi+\mathfrak{R}\,,
\end{align}
%\end{subequations}
where we have denoted the massless component by $\varphi = P_0 \Phi$ and the massive component by $\mathfrak{R}= \overline{P}_0 \Phi$. Splitting also the variation as $\delta \Phi = \delta\varphi+\delta\mathfrak{R}$, we obtain the equations of motion $\text{EOM}_\varphi$ and $\text{EOM}_\mathfrak{R}$ for $\varphi$ and $\mathfrak{R}$
\begin{subequations}
\begin{align}
\text{EOM}_\varphi (\varphi,\mathfrak{R}) &=P_0\text{EOM} (\varphi+\mathfrak{R})=-\eta_0 Q \varphi +P_0\mathcal{J}(\varphi+\mathfrak{R})\,,\label{eq:EOMphiH}\\
\text{EOM}_\mathfrak{R} (\varphi,\mathfrak{R}) &=\overline{P}_0\text{EOM} (\varphi+\mathfrak{R})=-\eta_0 Q \mathfrak{R} +\overline{P}_0\mathcal{J}(\varphi+\mathfrak{R})\,.
\end{align}
\end{subequations}

\subsubsection{Integrating out the massive fields}\label{sec221}
We want to integrate out the massive part $\mathfrak{R}$ of the string field $\Phi$ by solving the associated equation of motion $\text{EOM}_{\mathfrak{R}}$ for $\mathfrak{R}$ in terms of $\varphi$ and then substituting $\mathfrak{R}(\varphi)$ back into the full SFT action \eqref{eq:Shet}. In order to do so, we will need to fix the $Q$- and $\eta$-gauge symmetries for the massive part of the string field. To this end, let us  introduce the projectors $\Pi_Q = b_0^+ c_0^+$ and $\Pi_\eta = \xi_0 \eta_0$, together with $\overline{\Pi}_Q = 1-\Pi_Q = c_0^+ b_0^+$ and $\overline{\Pi}_\eta = 1-\Pi_\eta = \eta_0 \xi_0$.
%Note that we have
%\begin{align}
%\frac{b_0^+}{L_0^+}\overline{P}_0 Q +Q\frac{b_0^+}{L_0^+}\overline{P}_0 = \overline{P}_0\,,
%\end{align}
%where we have used that $[Q,b_0^+]=L_0^+$.

Let us first  fix the $\eta$-gauge symmetry of $\mathfrak{R}$. We will do so by splitting
\begin{align}
\mathfrak{R} = \mathcal{R}+\tilde{\mathcal{R}}\,,
\end{align}
where we define $\mathcal{R}=\Pi_\eta \mathfrak{R}$ and $\tilde{\mathcal{R}}=\overline{\Pi}_\eta \mathfrak{R}$. We will then fix the $\xi_0\mathfrak{R}=0$ gauge by setting $\tilde{\mathcal{R}}=0$. Doing so, the equation of motion $\text{EOM}_{\mathfrak{R}}$ gets split into two components (here we use that ${\rm bpz}(\Pi_\eta) = \overline{\Pi}_\eta$ and we substitute the gauge-fixed string field $\mathfrak{R}=\mathcal{R}$)
\begin{subequations}
\begin{align}
\text{EOM}_\mathcal{R}(\varphi,\mathcal{R}) &= \overline{\Pi}_\eta \text{EOM}_{\mathfrak{R}}(\varphi,\mathcal{R}) =- \eta_0 Q\mathcal{R} + \overline{\Pi}_\eta \overline{P}_0 \mathcal{J}(\varphi+\mathcal{R}),\label{eq:EOMRQH}\\
\text{EOM}_{\tilde{\mathcal{R}}}(\varphi,\mathcal{R}) &= {\Pi}_\eta \text{EOM}_{\mathfrak{R}}(\varphi,\mathcal{R}) = {\Pi}_\eta \overline{P}_0 \mathcal{J}(\varphi+\mathcal{R})\,.\label{eq:EOMRtH}
\end{align}
\end{subequations}
These are the equations of motion corresponding to the massive fields $\mathcal{R}$ , when these fields are partially gauge fixed by $\xi_0\mathcal{R}=0$.
If we now substitute $\Phi=\varphi+\mathcal{R}$ back into the action \eqref{eq:Shet} we obtain a partially gauge-fixed action
\begin{align}
S_\eta(\varphi,\mathcal{R}) = \frac{1}{2}\big\langle \eta_0\varphi,Q\varphi\big\rangle+\frac{1}{2}\big\langle \eta_0 \mathcal{R},Q\mathcal{R}\big\rangle +\mathcal{I}(\varphi+\mathcal{R})\,,\label{eq:ShetGF}
\end{align}
so that varying \eqref{eq:ShetGF} with respect to $\varphi$ and $\mathcal{R}$ will yield \eqref{eq:EOMphiH} (with $\mathfrak{R}=\mathcal{R}$) and \eqref{eq:EOMRQH}, respectively.
Notice however that in this way we miss the equation of motion \eqref{eq:EOMRtH}, which should then be interpreted as a gauge constraint (out-of-$\xi_0$ gauge equation).

Let us now proceed to fix the $Q$-gauge symmetry for the remaining component $\mathcal{R}$ of $\overline{P}_0 \Phi$. To this end, we  can decompose \cite{Asano:2016rxi}
\begin{align}
Q = c_0^+ L_0^+ + b_0^+ \mathcal{M}^+ +\widehat{{Q}}\,,
\end{align}
where we define $\mathcal{M}^+ = M^++\overline{M}^+$ and $\widehat{Q}=\widehat{Q}_B+\overline{\widehat{Q}}_B$ with
\begin{subequations}
\begin{align}
M^+&=-2\sum_{n>0} n c_{-n} c_{n}-2\sum_{q\geqslant \frac{1}{2}}\gamma_{-q}\gamma_{q}\,,\\
\overline{M}^+&=-2\sum_{n>0} m \overline{c}_{-n} \overline{c}_{n}
\end{align}
\end{subequations}
and, using conventions where $\gamma = e^{\phi}\eta$, $\beta = e^{-\phi}\p \xi $,
\begingroup\allowdisplaybreaks
\begin{subequations}
\begin{align}
\widehat{Q}_B&=\sum_{n\neq 0} c_{-n} L_n^\text{m}-\frac{1}{2}\sum_{\substack{
m,n\neq 0\\
m+n\neq 0
}}(m-n) \,:\!c_{-m}c_{-n}b_{m+n}\!:-\sum_{q+\frac{1}{2}\in\mathbb{Z}}\gamma_{-q}G_q^\text{m}+\nonumber\\
&\hspace{4cm}-\sum_{p+q\in\mathbb{Z}_{\neq 0}}\gamma_{-p}\gamma_{-q}b_{p+q}+\frac{1}{2}\sum_{\substack{m\in\mathbb{Z}_{\neq 0}\\ q+\frac{1}{2}\in\mathbb{Z}}}(m-2q)c_{-m}\gamma_{-q}\beta_{m+q}\,,\\
\overline{\widehat{Q}}_B &=\sum_{n\neq 0} \overline{c}_{-n} \overline{L}_n^\text{m}-\frac{1}{2}\sum_{\substack{
m,n\neq 0\\
m+n\neq 0
}}(m-n) \,:\!\overline{c}_{-m}\overline{c}_{-n}\overline{b}_{m+n}\!:\,.
\end{align}
\end{subequations}
\endgroup
Note that $\mathcal{M}^+$ and $\widehat{{Q}}$ do not contain $b_0$, $c_0$, $\overline{b}_0$, $\overline{c}_0$.
We then decompose
\begin{align}
\mathcal{R}=R+\tilde{R}\,,
\end{align}
where we denote $R=\Pi_Q\mathcal{R}$ and $\tilde{R}=\overline{\Pi}_Q\mathcal{R}$. The remaining portion of gauge symmetry, associated with $Q$, can then be fixed by setting $\tilde{R}=0$, that is imposing Siegel gauge $b_0^+\mathcal{R}=0$. The equation of motion $\text{EOM}_{\mathcal{R}}$ then splits into two components
\begin{subequations}
\begin{align}
\text{EOM}_R(\varphi,R) &= \Pi_Q \text{EOM}_{\mathcal{R}}(\varphi,R) = -\eta_0 c_0^+ L_0^+ R +\overline{\Pi}_Q\overline{\Pi}_\eta \overline{P}_0\mathcal{J}(\varphi+R),\label{eq:EOMRhete}\\
\text{EOM}_{\tilde{R}}(\varphi,R) &= \Pi_Q \text{EOM}_{\mathcal{R}}(\varphi,R) = -b_0 c_0\eta_0 \widehat Q R +{\Pi}_Q\overline{\Pi}_\eta \overline{P}_0\mathcal{J}(\varphi+R)\,.\label{eq:EOMRtQ}
\end{align}
\end{subequations}
Substituting $\mathcal{R} = R$ into $S_\eta(\varphi,\mathcal{R})$ \eqref{eq:ShetGF}, we obtain the action
\begin{align}
S_\text{gf}(\varphi,R) = \frac{1}{2}\big\langle \eta_0\varphi, Q\varphi\big\rangle+ \frac{1}{2}\big\langle \eta_0 R,c_0^+ L_0^+ R \big\rangle +\mathcal{I}(\varphi+R),
\end{align}
where all of the gauge symmetry of $\overline{P}_0\Phi$ has been fixed. Varying $S_\text{gf}(\varphi,R)$ with respect to $\varphi$ and $R$, we obtain the massless equation of motion \eqref{eq:EOMphiH} (where we substitute $\mathfrak{R}=R$) together with the gauge-fixed massive equation of motion \eqref{eq:EOMRhete}. Furthermore, we need to remember that on top of this, we still have to satisfy the two gauge constraints \eqref{eq:EOMRtH} (with $\mathcal{R}=R$) and \eqref{eq:EOMRtQ}. All of these represent a set of equations for $\varphi$ and $R$.
%Note that had we gauge-fixed both the $\eta$- and $Q$-symmetry simultaneously by decomposing the string field as
%\begin{align}
%\Phi = P_0\Phi +\overline{P}_0 \Pi_Q\Pi_\eta \Phi+\overline{P}_0 \Pi_Q\overline{\Pi}_\eta \Phi+\overline{P}_0 \overline{\Pi}_Q\Pi_\eta \Phi+\overline{P}_0 \overline{\Pi}_Q\overline{\Pi}_\eta \Phi\,,
%\end{align}
%we would have obtained three gauge constraints. However, it is easy to see that these are completely equivalent to the two gauge constraints \eqref{eq:EOMRtH} and \eqref{eq:EOMRtQ} because the additional gauge constraint comes simply from splitting \eqref{eq:EOMRtH} into two orthogonal subspaces by acting with $\Pi_Q$ and $\overline{\Pi}_Q$.
The massive gauge-fixed equation of motion \eqref{eq:EOMRhete} can be consistently solved to yield $R$ as a function of $\varphi$. Substituting $R(\varphi)$ into the massive gauge constraints \eqref{eq:EOMRtH} and \eqref{eq:EOMRtQ}, we will then show that these are automatically solved once we also assume that the massless equation of motion $\text{EOM}_\varphi$ holds. This will allow us to write down the effective action for $\varphi$ by simply substituting $R(\varphi)$ into $S_\text{gf}(\varphi,R)$, to finally get
\be
S_\text{eff}(\varphi)=S_\text{gf}(\varphi,R(\varphi)).
\ee

Let us first discuss solving the gauge-fixed massive equation of motion
\begin{align}
 \eta_0 c_0^+ L_0^+ R =c_0^+ b_0^+ \eta_0 \xi_0 \overline{P}_0\mathcal{J}(\varphi+R)\,.
\end{align}
Noting that $b_0^+ R= \xi_0 R=0$, we can first cancel both $c_0^+$ and $\eta_0$, then we can safely invert $L_0^+$ thanks to the projection outside its kernel $\overline{P}_0$. Altogether, we conclude we can equivalently write the massive EOM as
\begin{align}
R(\varphi) &=\frac{b_0^+}{L_0^+}\overline{P}_0\xi_0\mathcal{J}(\varphi+R(\varphi))\,.
\label{eq:RRecHet}
\end{align}
In order to lighten a bit the notation, let us introduce the products
\begin{align}
p_k(A_1,\ldots,A_k) \equiv \frac{b_0^+}{L_0^+}\overline{P}_0\xi_0[A_1,\ldots,A_k]\,,
\end{align}
together with the operator
\begin{align}
\mathcal{G}(\Phi) = \frac{b_0^+}{L_0^+}\overline{P}_0\xi_0\mathcal{J}(\Phi)\,.
\end{align}
The operator $\mathcal{G}$ can be expanded in terms of $p_k$ as
\begin{align}
\mathcal{G}(\Phi) &= \frac{1}{2!}p_2(\eta_0 \Phi,Q\Phi)+ \frac{1}{3!}\bigg(p_3(\eta_0 \Phi,Q\Phi,Q\Phi)+\nonumber\\
&\hspace{1.0cm}+p_2(\eta_0\Phi,[\Phi,Q\Phi])-\frac{1}{2}p_2(\Phi,[\Phi,\eta_0 Q\Phi])-\frac{1}{2}p_2(\Phi,[\eta_0 \Phi,Q\Phi])\bigg)+\mathcal{O}(\Phi^4)\,.
\end{align}
Then \eqref{eq:RRecHet} can be rewritten as
\begin{align}
R(\varphi) = \mathcal{G}(\varphi+R(\varphi))\,.
\end{align}
To solve this equation for $R$ we will assume  that $R(\varphi)$ fluctuates only in response to the fluctuations of $\varphi$. That is, we will put $R(0)=0$. Note that by doing this we will miss a number of classical solutions of the full SFT equation of motion where the massive sector is allowed to condense independently of the massless sector. Nevertheless, it completely covers the case at hand, namely the perturbative dynamics of the massless modes. We can then write the solution for $R(\varphi)$ in the form
\begin{align}
R(\varphi)=\mathcal{G}(\varphi+\mathcal{G}(\varphi+\mathcal{G}(\varphi+\ldots))).
\end{align}
It is easy to find that up to cubic order in $\varphi$ we have
\begingroup\allowdisplaybreaks
\begin{align}
R(\varphi) &= \frac{1}{2!}p_2(\eta_0 \varphi,Q\varphi)+\frac{1}{3!}\bigg(p_3(\eta_0 \varphi,Q\varphi,Q\varphi)+\nonumber\\
&\hspace{0.7cm}+p_2(\eta_0\varphi,[\varphi,Q\varphi])-\frac{1}{2}p_2(\varphi,[\varphi,\eta_0 Q\varphi])-\frac{1}{2}p_2(\varphi,[\eta_0 \varphi,Q\varphi])\bigg)+\nonumber\\
&\hspace{1.5cm}+\frac{1}{(2!)^2}p_2(\eta_0p_2(\eta_0\varphi,Q\varphi),Q\varphi)+\frac{1}{(2!)^2}p_2(\eta_0 \varphi,Qp_2(\eta_0\varphi,Q\varphi))+\mathcal{O}(\varphi^4)\,.\label{eq:Rvarphi}
\end{align}
\endgroup

The effective action for the massless excitations $\varphi$ is then
\begin{align}
S_\text{eff}(\varphi) = S_\text{gf}(\varphi,R(\varphi)) = \frac{1}{2!}\big\langle \eta_0\varphi, Q\varphi\big\rangle+ \frac{1}{2}\big\langle \eta_0 R(\varphi),c_0^+ L_0^+ R(\varphi) \big\rangle +\mathcal{I}(\varphi+R(\varphi))\,,
\label{eq:SeffHetGen}
\end{align}
%Note that it is straightforward to reproduce the above-presented large Hilbert space gauge-fixing procedure for the WZW-like open superstring field theory, both with or without an on-shell closed-string background being turned on (see \cite{Maccaferri:2019xx}).
and expanding this up to quartic order, we finally obtain
\begin{align}\label{seff}
S_\text{eff}(\varphi)&=\frac{1}{2!}\big\langle \eta_0\varphi, Q\varphi\big\rangle+\frac{1}{3!}\big\langle\eta_0 \varphi,[\varphi,Q\varphi]\big\rangle+\frac{1}{4!}\bigg(\big\langle \eta_0 \varphi,[\varphi,Q\varphi,Q\varphi]\big\rangle+\big\langle\eta_0 \varphi,[\varphi,[\varphi,Q\varphi]]\big\rangle\bigg)\nonumber\\
&\hspace{2cm}-\frac{1}{8}\bigg\langle[\eta_0\varphi,Q\varphi] , \xi_0\frac{b_0^+}{L_0^+}[\eta_0\varphi,Q\varphi]\bigg\rangle+O(\varphi^5).
\end{align}
But now it is important to understand what happens to the gauge constraints \eqref{eq:EOMRtH} and \eqref{eq:EOMRtQ} that we have to supplement in addition to the EOM derived from the effective action \eqref{seff}.  This analysis is carried out in appendix \ref{app:A} and the outcome is that the gauge constraints are satisfied (to the order we are interested in) thanks to the massless  equations (assuming the gauge-fixed massive equations have been already solved). So the gauge constraints  \eqref{eq:EOMRtH} and \eqref{eq:EOMRtQ} do not contain new physical information that is not already contained in \eqref{seff}. This is all we need.
 Interestingly this result can be generalized to all-orders in the case the interactions arrange themselves into cyclic $A_\infty$ or $L_\infty$ structures as it is the case for small Hilbert space theories, as it will be reported in \cite{Erbin:2019xx}.

%where we have noted that
%\begin{subequations}
%\begin{align}
%\frac{1}{3!}\big\langle\eta_0 R(\varphi),[\varphi,Q\varphi]\big\rangle&=\frac{1}{3!}\big\langle  R(\varphi),[\eta_0\varphi,Q\varphi]\big\rangle-\frac{1}{3!}\big\langle  R(\varphi),[\varphi,\eta_0 Q\varphi]\big\rangle\,,\\
%\frac{1}{3!}\big\langle\eta_0 \varphi,[R(\varphi),Q\varphi]\big\rangle&=
%\frac{1}{3!}\big\langle R(\varphi),[\eta_0 \varphi,Q\varphi]\big\rangle
%\,,\\
%\frac{1}{3!}\big\langle\eta_0 \varphi,[\varphi,QR(\varphi)]\big\rangle&=
%\frac{1}{3!}\big\langle R(\varphi),[\eta_0 \varphi,Q\varphi]\big\rangle+\frac{1}{3!}\big\langle R(\varphi),[\varphi,\eta_0 Q\varphi]\big\rangle
%\end{align}
%\end{subequations}

It is not difficult to check that varying the action \eqref{seff} (and using $\delta\varphi=P_0\delta\varphi$) we get back the equation of motion for the massless fields \eqref{eq:EOMphiH}
\be
\eta_0 Q \varphi =P_0\mathcal{J}(\varphi+R(\varphi)),\label{massless-eom}
\ee
where $R(\varphi)$ is given in \eqref{eq:Rvarphi}. If we find a solution to this equation we automatically have a solution to the full SFT since the massive equations have been
already solved and (importantly)  the $\eta$ and $Q$ gauge constraints are solved if the massless EOMs are solved, as shown in appendix \ref{app:A}.  Therefore, in the present perturbative setting, the massless EOM \eqref{massless-eom} is the only real requirement for the existence of a full SFT solution.\footnote{In the analogous open string setting of \cite{Vosmera:2019mzw, Mattiello:2019gxc} the  massless  equation was obtained in a
conceptually different approach: it was the obstruction to invert the kinetic operator in the full OSFT equation of motion  and thus the obstruction to the existence of a solution representing a true marginal boundary deformation.}

%

%%%%%%%%%%%%%%%%%%%%%%%%%%%%%%%
\section{Evaluation of the effective action}\label{sec3}
%%%%%%%%%%%%%%%%%%%%%%%%%%%%%%%
In this section we will analyze the various components of the effective action and we will compute the relevant physical couplings up to quartic order in
the massless fields. But before entering the details of the computation, we have to give a concrete look at the massless fields, i.e. the states $\varphi$ that we find in the kernel of $L_0^+$.

\subsection{The kernel of $L_0^+$}\label{ker}

In order to enumerate the field content of $\text{ker}\, L_0^{+}$ we will assume that the background which we are considering is associated with a unitary matter SCFT. That is, all matter states $\mathbb{V}_h$ have $h\geqslant 0$ and, in the matter CFT, the identity is the unique state with $h=0$. This assumption can be motivated for instance by working at zero-momentum in a compactification with unitary internal SCFT. Our computation will then produce algebraic couplings of the full effective action. %Unitarity of the matter SCFT will also turn out to be a crucial requirement for the localization of the quartic coupling to work (see  below for details).
We want to write down the most general state $\varphi$ which is level-matched
\begin{subequations}
\begin{align}
(b_0-\overline{b}_0)\varphi&=0\,,\\
(L_0-\overline{L}_0)\varphi&=0,
\end{align}
\end{subequations}
at total ghost number $+1$ and picture number $0$ such that $(L_0+\overline{L}_0)\varphi=0$, in the large Hilbert space. The result can be written as
\begin{align}
\varphi = \varphi_1+\varphi_2+\varphi_3\,,\label{kernel}
\end{align}
where
\begin{subequations}
\label{kernelL0}
\begin{align}
\varphi_1 &=\varphi_A+\varphi_D+\varphi_B \\
\varphi_2 &= \big(\p c+\overline{\p}\overline{c}\big)\Big(\xi c\mathbf{V}_{\frac12} e^{-\phi}+ c\xi \p \xi e^{-2\phi}\overline{c}\overline{\mathbf{V}}_1\Big)\,,\\
\varphi_3
&= c\mathbf{W}_1+\overline{c}\overline{\mathbf{W}}_1+ \eta e^{\phi}\mathbf{U}_\frac{1}{2}+C_1 c\p \phi +C_2\big(\p c+\overline{\p}\overline{c}\big)\,.
\end{align}
\end{subequations}
To start with $\varphi_1$ obeys the physical state condition $\eta_0 Q\varphi_1=0$. Its first component is \eqref{phi-phys}
\begin{align}
\varphi_A= \xi\, c \overline{c}\,\mathcal{V}_{\frac12,1}\,e^{-\phi}\,,
\end{align}
and it captures all of the physical massless fields at zero momentum except
\begin{align}
\varphi_D= D\, \xi\,Y\,Q(\p c-\overline{\p}\overline{c}),\quad\quad D\in\mathbb{R}
\end{align}
which is the zero-momentum ghost dilaton in the large Hilbert space. This state is universal and it is build using the inverse-picture changing operator
\be
Y(z)=c\del\xi e^{-2\phi}(z)=c\delta'(\gamma)(z),
\ee
which is a weight zero primary field in the cohomology of $Q$ with vanishing ghost number and picture $p=-1$. The ghost dilaton $\varphi_D$ is formally gauge-trivial but in fact it is not, because $(\p c-\overline{\p}\overline{c})$ is not an allowed closed-string state, see  \cite{Yang:2005ep} for a  discussion in the bosonic string. The remaining state in $\varphi_1$ is given by
\be
\varphi_B=B\,\xi Y Q(\p c+\overline{\p}\overline{c}),\quad B\in \mathbb{R}.
\ee
However differently from the ghost dilaton this state is gauge trivial in the semi-relative cohomology $b_0^-=0$.
\be
\varphi_B=B\Big(-Q(\,\xi Y (\p c+\overline{\p}\overline{c}))+\eta_0(\, \xi_0 X_0(Y(\p c+\overline{\p}\overline{c})))\Big).
\ee

The states $\varphi_2$ are Nakanishi-Lautrup-like (NL) fields and are spanned by $\mathbf{V}_{\frac12}$, $\overline{\mathbf{V}}_1$ which are generic holomorphic and anti-holomorphic matter conformal fields of weigth $(1/2,0)$ and $(0,1)$ respectively. These states are set to zero by the Siegel gauge condition $b_0^+=0$.

 The remainig fields $\varphi_3$ (spanned by the conformal fields $\mathbf{W}_1$, $\overline{\mathbf{W}}_1$, $\mathbf{U}_{\frac12}$ and the numbers $C_1,C_2$) are in the small Hilbert space $\eta_0=0$ and are therefore $\eta_0$-gauge-trivial
 \be
 \varphi_3=\eta_0(\xi\varphi_3).
 \ee

\subsection{Effective action for the physical fields}
Obviously we would like to focus on just the physical fields, however the possible elimination from the game of all the  fields in the kernel of $L_0^+$ except the physical ones $\varphi_A$ and $\varphi_D$ needs a careful analysis. We have addressed this problem in appendix \ref{app:B} and here we just briefly summarize the outcome.

First we partially gauge-fix the effective action \eqref{seff} by requiring $\xi_0\varphi=0$, that  setting $\varphi_3=0$ in \eqref{kernel}. In doing so we loose some of the massless equation of motion which become gauge constraints. But like in the massive case, these constraints are satisfied when the remaining massless fields solve the equation of motion.

Then we  integrate out $\varphi_2$ as a function of $\varphi_1$.  Here we importantly check that the possible generated new couplings from integrating out $\varphi_2$ do not contribute to the effective action up to quartic order.

Finally we
show that the pure gauge state $\varphi_B$ in $\varphi_1$ completely decouples.
The final outcome from appendix \ref{app:B} is that, up to quartic order, the correct effective action for the physical fields $\varphi_A$ and $\varphi_D$ is obtained from the
full effective action \eqref{seff} by simply setting $\varphi=\varphi_A+\varphi_D$
\begin{align}
S_{\text{eff,p}}(\varphi_A,\varphi_D)=S_{\text{eff}}(\varphi){\Big |}_{\varphi=\varphi_A+\varphi_D} \quad\text{(up to quartic order)}.
\end{align}
Therefore the effective action only depends on the physical fields $\varphi_A$ and $\varphi_D$.
From the decoupling of the pure gauge mode $\varphi_B$ shown in appendix \ref{app:B}, one may be tempted to conclude that also the similarly-looking
ghost-dilaton $\varphi_D$ decouples, observing that we can formally write
\begin{align}
\varphi_D = \xi_0 Q\chi_D\,
\end{align}
with $$\chi_D=2D\, Y c_0^-|0\rangle.$$ However $b_0^{-}\chi_D\neq 0$. As a consequence, we may not use the  computations in appendix \ref{app:B} to show that also $\varphi_D$ drops out from the quartic part of the action. It is however true that the computation \eqref{eq:WPStart} -- \eqref{eq:weakProj} shows that $\varphi_D$ in fact drops from the cubic part of the action. Therefore, we can write the effective action for physical fields only as
\begingroup\allowdisplaybreaks
\begin{align}
S_{\text{eff,p}}(\varphi_A,\varphi_D) &= \frac{1}{3!}\big\langle\eta_0 \varphi_A,[\varphi_A,Q\varphi_A]\big\rangle+\nonumber\\
&\hspace{0.5cm}+\bigg\{\frac{1}{4!}\bigg(\big\langle \eta_0 \varphi_\text{p},[\varphi_\text{p},Q\varphi_\text{p},Q\varphi_\text{p}]\big\rangle+\big\langle\eta_0 \varphi_\text{p},[\varphi_\text{p},[\varphi_\text{p},Q\varphi_\text{p}]]\big\rangle\bigg)+\nonumber\\
&\hspace{1cm}
+\frac{1}{8}\bigg\langle [\eta_0 \varphi_\text{p},Q\varphi_\text{p}],\frac{b_0^+}{L_0^+}\xi_0 \overline{P}_0[\eta_0 \varphi_\text{p},Q\varphi_\text{p}]\bigg\rangle+\mathcal{O}(\varphi_\text{p}^5)\bigg\}\bigg|_{\varphi_\text{p}=\varphi_A+\varphi_D}\,,
\end{align}
\endgroup
or, in the small Hilbert space ($\psi=\eta_0\varphi$)
\begingroup\allowdisplaybreaks
\begin{align}
S_{\text{eff,p},S}(\psi_A,\psi_D) &= \frac{1}{3!}\big\langle\psi_A,[\psi_A,X_0\psi_A]\big\rangle_S+\nonumber\\
&\hspace{0.5cm}+\bigg\{\frac{1}{4!}\bigg(\big\langle \psi_\text{p},[ \psi_\text{p},X_0\psi_\text{p},X_0 \psi_\text{p}]\big\rangle_S+\nonumber\\
&\hspace{2cm}+\big\langle\psi_\text{p},[\xi_0\psi_\text{p},[\psi_\text{p},X_0 \psi_\text{p}]+[\psi_\text{p},[\xi_0\psi_\text{p},X_0 \psi_\text{p}]]\big\rangle_S\bigg)+\nonumber\\
&\hspace{1cm}
-\frac{1}{8}\bigg\langle [\psi_\text{p},X_0 \psi_\text{p}],\frac{b_0^+}{L_0^+} \overline{P}_0[\psi_\text{p},X_0 \psi_\text{p}]\bigg\rangle_S+\mathcal{O}(\psi_\text{p}^5)\bigg\}\bigg|_{\psi_\text{p}=\psi_A+\psi_D}\,.
\end{align}
\endgroup
An important question at this point is what to do with the ghost dilaton $\varphi_D$ in the quartic part of the effective action.
%Assuming that the discussion in \cite{Bergman:1994qq}  in the context of the bosonic string  naturally generalizes to the superstring, then we would expect $\varphi_D$ to have vanishing effective potential. Moreover since a shift in the ghost dilaton is expected to correspond to a shift in the string coupling constant, and since If so  then we would expect that, even if not gauge-trivial, $\varphi_D$ should still decouple from the full effective action\footnote{See  \cite{Yang:2005ep} for a numerical check of this in bosonic closed string field theory.}, just as it happens for the cubic couplings
%\be
%S_{\text{eff, p}}(\varphi_A,\varphi_D)\mathrel{\stackon[2pt]{$=$}{$\scriptstyle?$}} S_{\text{eff, p}}(\varphi_A, 0)\quad {\rm (Dilaton\, Theorem?)}.\label{gd}
%\ee
%
This field is independent of the matter sector and therefore it is insensitive to the localization mechanism we discuss next. In the following we focus on the cubic and quartic couplings of $\varphi_A$ alone and we will simply set the ghost dilaton to zero $\varphi_D=0$.
The computation of the ghost dilaton couplings is however a quite important question that should be addressed, using appropriate  tools which are however beyond the scope of this paper. See \cite{Bergman:1994qq} and \cite{Yang:2005ep} for related results in the context of bosonic closed string field theory. %
%In appendix \ref{app:B} we show how we can rid of all of the fields in the kernel of $L_0^+$ except the physical ones $\varphi_A$ and $\varphi_D$.
%Therefore  the non-trivial fields in ker$(L_0^+)$ are $\varphi_A$, $\varphi_D$ which are both physical and the NL fields $\varphi_2$.
%\be
%S_{\rm eff}=S_{\rm eff}(\varphi_A,\varphi_D;\varphi_2).
%\ee
%{\bf From here on it is not clear what to say and what not to say}
%
%In general (differently from $\varphi_3$) the Nakanishi-Lautrup-like field $\varphi_2$ cannot be simply removed by fixing Siegel gauge without generating non-trivial gauge constraints. Therefore, $\varphi_2$ should be integrated out in order to obtain an effective action for $\varphi_1$ only, potentially generating new couplings of $\varphi_1$ along the way.
%
%Finally, given the effective action for physical fields $\varphi_1$, one still needs to address the couplings involving one or more insertions of the ghost dilaton vertex operator.
\subsection{$\N=2$ decomposition}
If we ignore the ghost dilaton, we remain with the physical field $\varphi_A$,
which has the general form
\begin{equation}\label{phiA}
	\varphi_A(z, \overline z)
		= \xi(z) \, c \overline c \VVV_{\frac12,1}(z, \overline z) \, \e^{-\phi(z)},
	\qquad
	\VVV_{\frac{1}{2},1}(z, \overline z)
		= \varepsilon_{ik} \, \VV^i_{\frac12}(z) \, \overline\WW^k_1(\overline z)
\end{equation}
and satisfies the free equation of motion
\begin{equation}
	\label{eq:phi-eom}
	\eta_0 Q \varphi_A
		= 0.
\end{equation}

% The cubic term can be rewritten as
% \begin{equation}
% 	S_3
% 		= \aver{\eta_0 \varphi, P_0 [ \varphi, Q \varphi ]}
% 		= \aver{\varphi, P_0 [ \eta_0 \varphi, Q \varphi ]}.
% \end{equation}

We now assume that the holomorphic side of the worldsheet has a global $\mc N = 2$  supersymmetry in the matter sector and that  the superconformal primary  $\V_{1/2}$ can be written as the sum of two fields of opposite $R$ charge
\begin{equation}
	\V_{\frac12}
		= \V_{\frac12}^+ + \V_{\frac12}^-\,.
\end{equation}
Denoting $\mathbb{V}_1 = G_{-1/2}\mathbb{V}_{1/2}$, we then have
\begin{equation}
\mathbb{V}_1 = \V_{1}^+ + \V_{1}^-\,,
\end{equation}
where $\V_{1}^+=G_{-1/2}^-\V_{1/2}^+$, $\V_{1}^-=G_{-1/2}^+\V_{1/2}^-$, together with $G_{-1/2}^+\V_{1/2}^+=0$, $G_{-1/2}^-\V_{1/2}^-=0$.The fields $\V_{\frac12}^\pm$ are charged under the $\mathrm{U}(1)$ $R$-symmetry generated by $J_0$ while the fields $\V_{1}^\pm$ are neutral:
\begin{equation}
	J_0 \V_{\frac12}^\pm
		= \pm \V_{\frac12}^\pm,
	\qquad
	J_0 \V_{1}^\pm
		= 0.
\end{equation}
Correlation functions are non-zero only if the total $R$-charge is zero.
In the same manner, we also decompose the total massless field as
\begin{equation}\label{decomp}
	\varphi_A
		= \varphi^+ + \varphi^-,
		\qquad
	\varphi^\pm
		= \xi c \overline c \VVV_{1/2,1}^\pm \e^{-\phi}.
\end{equation}
Acting with $Q$ gives a combination of the charged and neutral fields, while acting with $\eta$ does not change the charge:
\begin{equation}
	Q \varphi^\pm
		= c \overline c \VVV_{1,1}^\pm - \eta \e^{\phi} \overline c \VVV_{1/2,1}^\pm,
	\qquad
	\eta_0 \varphi^\pm
		= c \overline c \VVV_{1/2,1}^\pm \e^{-\phi}.
\end{equation}
We see that there will be a tension between conservations of ghost number and $R$-charge for terms involving $Q\varphi$: the first term has $N_{\text{gh}}^{bc} = 1$ and $J_0 = 0$, while the second has $N_{\text{gh}}^{bc} = 0$ and $J_0 = 1$.
In total, only the terms such that $N_{\text{gh}}^{bc} = 6$ and $J_0 = 0$ can contribute.
%%%%%%%%%%%%%%%%%%%%
\subsection{Cubic couplings}
%%%%%%%%%%%%%%%%%%%%

Given the explicit representation \eqref{phiA} we can directly compute the cubic coupling of the action.
Using $\varphi_A=P_0\varphi_A$ and  \eqref{eq:weakProj} we have that\footnote{We normalize the ghost correlators so that
\begin{align}
\big\langle 0\big| \xi_0  c_{-1} \overline{c}_{-1} c_0 \overline{c}_0 c_1 \overline{c}_1 e^{-2\phi}
\big|0\big\rangle=1\,.
\end{align}
}

\begingroup\allowdisplaybreaks
\begin{subequations}
\begin{align}
\frac{1}{3!}\big\langle \eta_0\varphi_A,[\varphi_A,Q\varphi_A]\big\rangle
%&=\frac{2}{3!}\varepsilon_{il}\varepsilon_{jm}\varepsilon_{kn}\big\langle 0\big|  \overline{c}_{-1}e^{-\phi} c_{-1}\xi_0 c_0^- c_0^+ c_1 \overline{c}_1 e^{-\phi}
%\big|0\big\rangle\times\nonumber\\
%&\hspace{3cm}
%\times\big\langle \mathbb{V}^i_\frac{1}{2}\big|\{\mathbb{V}^{j}_\frac{1}{2}\mathbb{V}_1^k\}_1\big\rangle
%\big\langle \overline{\mathbb{W}}_1^l\big|\{\overline{\mathbb{W}}_1^m\overline{\mathbb{W}}_1^n\}_1 \big\rangle \\
&=\frac{1}{3!}\varepsilon_{il}\varepsilon_{jm}\varepsilon_{kn}\big\langle 0\big| \xi_0  c_{-1} c_0 c_1\overline{c}_{-1} \overline{c}_0\overline{c}_1 e^{-2\phi}
\big|0\big\rangle\times\nonumber\\
&\hspace{3cm}
\times\big\langle \mathbb{V}^i_\frac{1}{2}\big|\{\mathbb{V}^{j}_\frac{1}{2}\mathbb{V}_1^k\}_1\big\rangle
\big\langle \overline{\mathbb{W}}_1^l\big|\{\overline{\mathbb{W}}_1^m\overline{\mathbb{W}}_1^n\}_1 \big\rangle \\
&=- \frac{1}{3!}\varepsilon_{il}\varepsilon_{jm}\varepsilon_{kn}
\big\langle \mathbb{V}^i_\frac{1}{2}\big|\{\mathbb{V}^{j}_\frac{1}{2}(G_{-\frac{1}{2}}\mathbb{V}_\frac{1}{2}^k)\}_1\big\rangle\big\langle\overline{\mathbb{W}}_1^l\big|\{\overline{\mathbb{W}}_1^m\overline{\mathbb{W}}_1^n\}_1 \big\rangle\,.
\end{align}
\end{subequations}
\endgroup
This is the universal form of the cubic coupling, which is given in terms of the three-point functions in the matter sector. However, whenever the $\mathcal{N}=2$ decomposition is possible, that is whenever we can write $\mathbb{V}_\frac{1}{2}^i =(\mathbb{V}_\frac{1}{2}^i)^++(\mathbb{V}_\frac{1}{2}^i)^- $, we in fact have
\begin{align}
\big\langle \mathbb{V}^i_\frac{1}{2}\big|\big\{\mathbb{V}^{j}_\frac{1}{2}(G_{-\frac{1}{2}}\mathbb{V}_\frac{1}{2}^k)\big\}_1\big\rangle =0\label{eq:projTest}
\end{align}
for all $i,j,k$.
The  proof of this claim, which proceeds along the lines of \cite{Vosmera:2019mzw},
is presented in appendix \ref{app:C}. Therefore, given \eqref{decomp}, the cubic potential vanishes
\begin{align}
\frac{1}{3!}\big\langle \eta_0\varphi_A,[\varphi_A,Q\varphi_A]\big\rangle{\Big|}_{\varphi_A=\varphi^++\varphi^-}=0.\label{no-cubic}
\end{align}
Note that \eqref{eq:projTest} actually implies a somewhat stronger result, namely that
\begin{align}
P_0[\eta_0 \varphi_A,Q\varphi_A]=0\label{eq:ProjCon},
\end{align}
whenever  $\varphi_A=\varphi^++\varphi^-$. Expanding the equation of motion following from the effective action $S_\text{eff}(\varphi_A)$ order by order in $\varphi_A$, one finds that \eqref{eq:ProjCon} is precisely the equation of motion arising at second order. Hence, we can conclude that to second order, we automatically satisfy the condition for existence of a marginal deformation by $\varphi_A$ whenever we have an $\mathcal{N}=2$ decomposition of $\varphi_A$.

\subsection{Localization of the quartic effective potential}

Now we finally  focus on the quartic interaction from \eqref{seff}, which can be decomposed into a vertex interaction and an interaction with a propagator
\begin{subequations}
\begin{align}
	S_{\text{eff}, 4}
		&
		= S_{\text{eff},4}^{\text{prop}} + S_{4}^{\times},
	\\
	\label{seff-prop}
	S_{\text{eff},4}^{\text{prop}}
		&
		= - \frac{1}{8} \aver{\overline P_0 [\eta_0 \varphi_A, Q \varphi_A ], \xi_0 \frac{b_0^+}{L_0^+} [\eta_0 \varphi_A, Q \varphi_A]},
	\\
	\label{seff-cont}
	S_{4}^{\times}
		&
		= \frac{1}{4!} \aver{\eta_0 \varphi, \big[\varphi_A, [\varphi_A, Q \varphi_A]\big]}
			+ \frac{1}{4!} \aver{\eta_0 \varphi_A, [ \varphi_A, (Q \varphi_A)^2]}.
\end{align}
\end{subequations}
%First, we show how it localizes when taking into account that the massless field $\varphi$ is a short $\mathcal N = 2$ multiplet.
%Second, we rewrite the localized potential in terms of auxiliary fields.
%

We now proceed with the explicit evaluation of the above quantities, assuming the $\N=2$ decomposition of $\varphi_A$ \eqref{decomp}.\footnote{Given the formal similarity of heterotic string field theory with open superstring field theory with stubs, it is not difficult to adapt the original computation  of \cite{Vosmera:2019mzw} to here (and this is in fact how we first arrived to the result). Here, however, we will  follow analogous steps as in the original paper \cite{Maccaferri:2018vwo}.}

% TODO: replace 1/2 → \frac12
% TODO: replace \varphi → \varphi_A

%We will need the ghost number of the structures appearing in the action:
%\begin{equation}
%	N_{\text{gh}}^{bc}([\cdot, \cdot])
%		= - 1,
%	\qquad
%	N_{\text{gh}}^{bc}([\cdot, \cdot, \cdot])
%		= - 3,
%	\qquad
%	N_{\text{gh}}^{bc}(\aver{\cdot, \cdot})
%		= 1.
%\end{equation}

\paragraph{Propagator terms \eqref{seff-prop}.}

The total ghost number of the propagator term
\begin{equation}
	S_{\text{eff},4}^{\text{prop}}
		= - \frac{1}{8} \aver{\overline P_0 [\eta_0 \varphi, Q \varphi ], \xi_0 \frac{b_0^+}{L_0^+} [\eta_0 \varphi, Q \varphi]}
\end{equation}
reads
\begin{equation}
	N_{\text{gh}}^{bc}
		= 2 + 2 N_{\text{gh}}^{bc}(Q \varphi)
\end{equation}
which requires $N_{\text{gh}}^{bc}(Q \varphi) = 2$: this selects the neutral component.
This means that the fields on which $\eta_0$ acts must have opposite charges to achieve $J_0 = 0$.
The non-vanishing terms are equal by pairs since $b_0^+/L_0^+$ is BPZ even and we find:
\begin{equation}
	\begin{aligned}
	S_{\text{eff},4}^{\text{prop}}
		=
			&
			- \frac{1}{4} \aver{\overline P_0 [\eta_0 \varphi^+, Q \varphi^+],
				\xi_0 \frac{b_0^+}{L_0^+} [\eta_0 \varphi^-, Q \varphi^+]}
			\\ &
			- \frac{1}{4} \aver{\overline P_0 [\eta_0 \varphi^+, Q \varphi^-],
				\xi_0 \frac{b_0^+}{L_0^+} [\eta_0 \varphi^-, Q \varphi^-]}
			\\ &
			- \frac{1}{4} \aver{\overline P_0 [\eta_0 \varphi^+, Q \varphi^+],
				\xi_0 \frac{b_0^+}{L_0^+} [\eta_0 \varphi^-, Q \varphi^-]}
			\\ &
			- \frac{1}{4} \aver{\overline P_0 [\eta_0 \varphi^+, Q \varphi^-],
				\xi_0 \frac{b_0^+}{L_0^+} [\eta_0 \varphi^-, Q \varphi^+]}.
	\end{aligned}
\end{equation}
Note that the first two terms are exchanged under $\varphi^\pm \leftrightarrow \varphi^\mp$, while the last two terms are invariant.

The next step is to show that, in fact, the first two terms vanish.
This can be seen by moving around the operators and showing that the result violates the charge or ghost number conservation.
For the first term, the idea is to move the $Q$ from the first bracket to act on $\varphi^-$ in the second bracket, in order to violate the charge conservation:
\begin{equation}
	\aver{\overline P_0 [\eta_0 \varphi^+, Q \varphi^+],
			\xi_0 \frac{b_0^+}{L_0^+} [\eta_0 \varphi^-, Q \varphi^+]}
		= \aver{\overline P_0 [\eta_0 \varphi^+, \varphi^+],
			\frac{b_0^+}{L_0^+} [Q \varphi^-, Q \varphi^+]}
		= 0.
\end{equation}
This vanishes because ghost number conservation picks the neutral terms in $Q \varphi^\pm$, but then $J_0 \neq 0$.
The second term vanishes exactly in the same way.

We can now analyze the two other terms to bring them in a simpler form.
The idea is to remove all propagators such that only terms with projectors and contact interactions remain.
This is achieved by moving again $Q$ such that it can hit the propagator.
For the 3rd term, we extract $Q$ and $\eta_0$ from both sides in a symmetric way.
Simplifying and using $R$-charge and ghost number conservations lead to:
\begin{equation}
	\begin{aligned}
	2 \aver{\overline P_0 [\eta_0 \varphi^+, Q \varphi^+],
			\xi_0 \frac{b_0^+}{L_0^+} [\eta_0 \varphi^-, Q \varphi^-]}
		=
			&
			- \aver{\overline P_0 [\eta_0 \varphi^-, \varphi^-], [\varphi^+, Q \varphi^+]}
			\\ &
			- \aver{\overline P_0 [\eta_0 \varphi^+, \varphi^+], [\varphi^-, Q \varphi^-]}.
		\end{aligned}
\end{equation}
In the 4th term, we use the relation
\begin{equation}
	[\eta_0 \varphi^+, Q \varphi^-]
		= - \eta_0 Q [\varphi^+, \varphi^-]
			- [Q \varphi^+, \eta_0 \varphi^-],
\end{equation}
in both sides symmetrically.
Again, simplifying and conservation laws leads to
\begin{equation}
	\begin{aligned}
	2 \aver{\overline P_0 [\eta_0 \varphi^+, Q \varphi^-],
			\xi_0 \frac{b_0^+}{L_0^+} [\eta_0 \varphi^-, Q \varphi^+]}
		=
			&
			- \aver{\overline P_0 [\varphi^-, \varphi^+], [\eta_0 \varphi^-, Q \varphi^+]}
			\\ &
			+ \aver{\overline P_0 [\varphi^-, \varphi^+], [\eta_0 \varphi^+, Q \varphi^-]}.
		\end{aligned}
\end{equation}

Putting the pieces back together, the action \eqref{seff-prop} becomes:
\begin{equation}
	\label{seff-prop-N2}
	\begin{aligned}
	S_{\text{eff},4}^{\text{prop}}
		=
			&\
			\frac{1}{8} \aver{\overline P_0 [\eta_0 \varphi^-, \varphi^-], [\varphi^+, Q \varphi^+]}
			+ \frac{1}{8} \aver{\overline P_0 [\eta_0 \varphi^+, \varphi^+], [\varphi^-, Q \varphi^-]}
			\\ &
			+ \frac{1}{8} \aver{\overline P_0 [\varphi^-, \varphi^+], [\eta_0 \varphi^-, Q \varphi^+]}
			- \frac{1}{8} \aver{\overline P_0 [\varphi^-, \varphi^+], [\eta_0 \varphi^+, Q \varphi^-]}.
	\end{aligned}
\end{equation}
This matches (A.7) in \cite{Maccaferri:2018vwo}.
Notice that this part of the action does not contain anymore propagators.
Writing $\overline P_0 = 1 - P_0$, it splits in \emph{localized} and \emph{contact} contributions:
\begin{equation}
	S_{\text{eff},4}^{\text{prop}}
		= S_{\text{eff},4}^{\text{loc}} + S_{\text{eff},4}^{\text{con}},
\end{equation}
where:
\begin{subequations}
\begin{align}
	\label{seff-prop-N2-loc}
	S_{\text{eff},4}^{\text{loc}}
		&
		= -\frac{1}{8} \aver{P_0 [\eta_0 \varphi^-, \varphi^-], [\varphi^+, Q \varphi^+]}
			- \frac{1}{8} \aver{P_0 [\eta_0 \varphi^+, \varphi^+], [\varphi^-, Q \varphi^-]}
			\\ & \qquad
		\nonumber
			- \frac{1}{8} \aver{P_0 [\varphi^-, \varphi^+], [\eta_0 \varphi^-, Q \varphi^+]}
			+ \frac{1}{8} \aver{P_0 [\varphi^-, \varphi^+], [\eta_0 \varphi^+, Q \varphi^-]},
	\\
	\label{seff-prop-N2-con}
	S_{\text{eff},4}^{\text{con}}
		&
		=  \frac{1}{8} \aver{[\eta_0 \varphi^-, \varphi^-], [\varphi^+, Q \varphi^+]}
			+\frac{1}{8} \aver{[\eta_0 \varphi^+, \varphi^+], [\varphi^-, Q \varphi^-]}
			\\ & \qquad
		\nonumber
			+ \frac{1}{8} \aver{[\varphi^-, \varphi^+], [\eta_0 \varphi^-, Q \varphi^+]}
			- \frac{1}{8} \aver{[\varphi^-, \varphi^+], [\eta_0 \varphi^+, Q \varphi^-]}.
\end{align}
\end{subequations}

We will now see, exactly as in \cite{Maccaferri:2018vwo}, the contact contribution will cancel with the elementary quartic vertex

\paragraph{Elementary quartic vertex \eqref{seff-cont}.}

The elementary quartic vertex is made of two terms:
\begin{equation}
	S_{4}^{\times}
		= \frac{1}{4!} \aver{\eta_0 \varphi, \big[\varphi, [\varphi, Q \varphi]\big]}
			+ \frac{1}{4!} \aver{\eta_0 \varphi, [ \varphi, (Q \varphi)^2]}.
\end{equation}
The ghost number of the first term (double $2$-product) is
\begin{equation}
	N_{\text{gh}}^{bc}
		= 5
				+ N_{\text{gh}}^{bc}(Q \varphi).
\end{equation}
This means that $Q \varphi^\pm$ picks the charged term, and there must be an equal number of $\varphi^\pm$ in the expression.
The ghost number of the second term ($3$-product) reads:
\begin{equation}
	N_{\text{gh}}^{bc}
		= 2 + 2 N_{\text{gh}}^{bc}(Q \varphi).
\end{equation}
It does not vanish only if $N_{\text{gh}}^{bc}(Q \varphi) = 2$, such that $Q \varphi^\pm$ picks the neutral term, and thus $\eta \varphi^\pm$ and $\varphi^\pm$ must have opposite signs.

Before proceeding, we display a set of relations for the term with the $3$-product which can be used to simplify further the expressions.
First, terms with opposite signs are in fact equal
\begin{equation}
	\aver{\eta_0 \varphi^-, [ \varphi^+, (Q \varphi)^2]}
		= \aver{\eta_0 \varphi^+, [ \varphi^-, (Q \varphi)^2]},
\end{equation}
as can be shown by using cyclicity, that $\eta_0$ is a derivative of the $3$-product and of the inner-product, and the equation of motion \eqref{eq:phi-eom}.
Moreover, using the $L_\infty$ relation
\begin{align*}
	0
		= Q [\varphi^-, \varphi^+, Q \varphi^+]
			&+ [Q \varphi^-, \varphi^+, Q \varphi^+]
			- [\varphi^-, Q \varphi^+, Q \varphi^+]
			\\ &
			- \big[ \varphi^-, [\varphi^+, Q \varphi^+] \big]
			+ \big[ \varphi^+, [\varphi^-, Q \varphi^+] \big]
			+ \big[ Q \varphi^+, [\varphi^-, \varphi^+] \big]
\end{align*}
together with conservation laws show that the terms with $(Q\varphi^\pm)^2$ vanish:
\begin{align*}
	\aver{\eta_0 \varphi^+, [ \varphi^-, Q \varphi^+, Q \varphi^+]}
		= \aver{\eta_0 \varphi^+, [ \varphi^-, Q \varphi^+, Q \varphi^+]}
		= 0.
\end{align*}
Using these relations, the action reads:
\begin{equation}
	\label{seff-cont-N2}
	\begin{aligned}
	S_4^{\times}
		=
			&
			- \frac{1}{4!} \aver{[\eta_0 \varphi^+, \varphi^+], [\varphi^-, Q \varphi^-]}
			- \frac{1}{4!} \aver{[\eta_0 \varphi^+, \varphi^-], [\varphi^+, Q \varphi^-]}
			\\ & \quad
			- \frac{1}{4!} \aver{[\eta_0 \varphi^+, \varphi^-], [\varphi^-, Q \varphi^+]}
			- \frac{1}{4!} \aver{[\eta_0 \varphi^-, \varphi^+], [\varphi^+, Q \varphi^-]}
			\\ & \quad
			- \frac{1}{4!} \aver{[\eta_0 \varphi^-, \varphi^+], [\varphi^-, Q \varphi^+]}
			- \frac{1}{4!} \aver{[\eta_0 \varphi^-, \varphi^-], [\varphi^+, Q \varphi^+]}
			\\ & \quad
			+ \frac{1}{6} \aver{\eta_0 \varphi^+, [ \varphi^-, Q \varphi^+, Q \varphi^-]}.
	\end{aligned}
\end{equation}

\paragraph{Rewriting the action}

The next step is to move $\eta_0$ and $Q$, to use the $L_\infty$ relation and to invoke conservation laws to write \eqref{seff-prop-N2-con} and \eqref{seff-cont-N2} in terms of independent quantities.
Ultimately, we find that only $4$ structures are independent:
\begin{equation}
	\begin{aligned}
	C_1
		&= \aver{[\varphi^-, \varphi^+], [\eta_0 \varphi^+, Q \varphi^-]},
	\qquad &
	C_2
		&= \aver{[\varphi^-, \varphi^+], [\eta_0 \varphi^-, Q \varphi^+]}
	\\
	C_3
		&= - \aver{[\eta_0 \varphi^+, \varphi^-], [\varphi^+, Q \varphi^-]},
	\qquad &
	D_{+-}
		&= \aver{\eta_0 \varphi^+, [ \varphi^-, Q \varphi^+, Q \varphi^-]}.
	\end{aligned}
\end{equation}

The $L_\infty$ relation
\begin{equation}
		\begin{aligned}
	0
		= Q [\varphi^i, \varphi^j, Q \varphi^k]
			&+ [Q \varphi^i, \varphi^j, Q \varphi^k]
			- [\varphi^i, Q \varphi^j, Q \varphi^k]
			\\ &
			- \big[ \varphi^i, [\varphi^j, Q \varphi^k] \big]
			+ \big[ \varphi^j, [\varphi^i, Q \varphi^k] \big]
			+ \big[ Q \varphi^k, [\varphi^i, \varphi^j] \big],
	\end{aligned}
\end{equation}
where $i, j, k = \pm$, can be used to rearrange terms where two fields with the same sign are contained in one bracket.
Note that the first term will never contribute because it contains $Q \eta_0 \varphi^\pm = 0$ after integrating by part.

After straightforward computations, we find that the different terms of the propagator part \eqref{seff-prop-N2-con} can be expressed as:
\begin{equation}
	\begin{aligned}
	\aver{[\eta_0 \varphi^-, \varphi^-], [\varphi^+, Q \varphi^+]}
		&
		= - D_{+-} - 2 C_2 - C_1 - C_3,
	\\
	\aver{[\eta_0 \varphi^+, \varphi^+], [\varphi^-, Q \varphi^-]}
		&
		= - D_{+-} - C_3 + C_1,
	\\
	\aver{[\varphi^-, \varphi^+], [\eta_0 \varphi^-, Q \varphi^+]}
		&
		= C_2,
	\\
	\aver{[\varphi^-, \varphi^+], [\eta_0 \varphi^+, Q \varphi^-]}
		&
		= C_1.
	\end{aligned}
\end{equation}
Summing all contributions, we find that \eqref{seff-prop-N2-con} reads:
\begin{equation}
	S_{\text{eff},4}^{\text{con}}
		= - \frac{1}{4!} (3 C_1 + 3 C_2 + 6 C_3 + 6 D_{+-}).\label{pip1}
\end{equation}
Similarly, the different terms of the contact interaction part \eqref{seff-cont-N2} can be written as:
\begin{equation}
	\begin{aligned}
	\aver{[\eta_0 \varphi^+, \varphi^+], [\varphi^-, Q \varphi^-]}
		&
		= - D_{+-} - C_3 + C_1,
	\\
	\aver{[\eta_0 \varphi^+, \varphi^-], [\varphi^+, Q \varphi^-]}
		&
		= - C_3,
	\\
	\aver{[\eta_0 \varphi^+, \varphi^-], [\varphi^-, Q \varphi^+]}
		&
		= - C_1 - C_3,
	\\
	\aver{[\eta_0 \varphi^-, \varphi^+], [\varphi^+, Q \varphi^-]}
		&
		= - C_1 - C_3,
	\\
	\aver{[\eta_0 \varphi^-, \varphi^+], [\varphi^-, Q \varphi^+]}
		&
		= - C_2 - C_1 - C_3,
	\\
	\aver{[\eta_0 \varphi^-, \varphi^-], [\varphi^+, Q \varphi^+]}
		&
		= - D_{+-} - 2 C_2 - C_1 - C_3.
	\end{aligned}
\end{equation}
Adding together all terms, we get:
\begin{equation}
	S_{4}^{\times}
		= \frac{1}{4!} (3 C_1 + 3 C_2 + 6 C_3 + 6 D_{+-}).\label{pip2}
\end{equation}
As a result, the contribution from the contact interaction \eqref{pip1} of the propagator term exactly cancels the vertex interaction \eqref{pip2}:
\begin{equation}
	S_{\text{eff},4}^{\text{con}}+ S_4^{\times}
		= 0.
\end{equation}
This implies that the quartic potential is given solely by the localized action \eqref{seff-prop-N2-loc}
\begin{equation}
	\begin{aligned}
	S_{\text{eff},4}
		= S_{\text{eff},4}^{\text{loc}}
		&
		=- \frac{1}{8} \aver{[\eta_0 \varphi^-, \varphi^-], P_0 [\varphi^+, Q \varphi^+]}
			- \frac{1}{8} \aver{[\eta_0 \varphi^+, \varphi^+], P_0 [\varphi^-, Q \varphi^-]}
			\\ & \qquad
			- \frac{1}{8} \aver{[\varphi^-, \varphi^+], P_0 [\eta_0 \varphi^-, Q \varphi^+]}
			- \frac{1}{8} \aver{[\varphi^+, \varphi^-], P_0 [\eta_0 \varphi^+, Q \varphi^-]}.
	\end{aligned}
\end{equation}
As described in the introduction, the effective action receives contribution only from the boundary of the moduli space.
%Indeed, the projector $P_0$ is equivalent to an infinite-length tube connecting the two $3$-point vertices given by $[\cdot, \cdot]$.

 %%%%%%%%%%%%%%%%%%%%%%
\subsection{Auxiliary fields}
%%%%%%%%%%%%%%%%%%%%%%
Now we would like to concretely analyze the states entering in the localized potential.
In order to do so we have to extract the Fock space components of states of the form $P_0[\Phi_1,\Phi_2]$ where
\be
\Phi_i=\Phi_i(0,0)|0\rangle_{SL(2,\CC)}\0
\ee
are primary states of weight $(0,0)$.
The 2-product $[\cdot,\cdot]$ is not uniquely defined as there are infinite choices that will be consistent with the $L_\infty$-relations (see for example \cite{Erler:2019loq}) however, from the symmetry of the three-vertex,  the product of two
conformal primaries of weight $(0,0)$ can be written in a form which naturally generalizes the open string star product as described in \cite{wedges}
\be
[\Phi_1,\Phi_2]=b_0^-\delta(L_0^-)\,e^{\sum_{n\geq1}v_n L_{-2n}+\overline v_n \overline L_{-2n}}\,\Phi_1(x,\overline x)\Phi_2(-x,-\overline x)|0\rangle_{SL(2,\CC)},
\ee
where the $v_n$'s and the location $(x,\overline x)$ depends on the chosen three-strings vertex.
From this general expression it is obvious that, provided that the OPE of $\Phi_1$ and $\Phi_2$ does not contain states with conformal dimension less than $-1$ in either left- or right-moving sector (this will always turn out to be true in our computations where we assume that the matter SCFT is unitary), the presence of $P_0$ will only select the identity component from the exponential and the OPE contribution of weight $(0,0)$ (if present). In such cases we therefore get the level $(0,0)$ Fock state
\be
P_0[\Phi_1,\Phi_2]=b_0^-\{\Phi_1\,\Phi_2\}_{0,0}(0,0)|0\rangle_{SL(2,\CC)},
\ee
where $\{\Phi_1,\Phi_2\}_{0,0}(z,\overline z)$ is the weight $(0,0)$ field which is found in the symmetric OPE of the weight zero fields  $\Phi_1$ and $\Phi_2$
\be
\Phi_1(z,\overline z)\Phi_2(-z,-\overline z)=\sum_{k,\bar k}(2z)^{-k}(2\overline z)^{-\bar k}\{\Phi_1\,\Phi_2\}_{k,\bar k}(0,0).
\ee
With this understanding we can easily compute the localized fields entering the effective quartic potential by standard OPE
\be
P_0[\varphi^\pm,Q\varphi^\pm]&=&c\overline c \,\HH^\pm_{1,1}(0,0)\,|0\rangle\\
P_0[\varphi^\pm,\eta_0\varphi^\pm]&=&-c\overline c(\del c+\overline\del\overline c)\,\xi e^{-2\phi} \,\HH^\pm_{1,1}(0,0)\,|0\rangle\\
P_0[\eta_0\varphi^\pm,Q\varphi^\mp]&=&\mp c\overline c\,\eta\,\HH_{0,1}(0,0)\,|0\rangle\\
P_0[\varphi^\pm,\varphi^\mp]&=&\pm c\overline c(\del c+\overline\del\overline c)\,\xi\del\xi e^{-2\phi}\, \HH_{0,1}(0,0)|0\rangle.
\ee
The heterotic auxiliary fields $\HH$ appearing above are matter primaries which can be  found in the OPEs of the physical matter fields
\be
\VVV^\pm_{\frac12,1}(z,\overline z)\VVV^\pm_{\frac12,1}(-z,-\overline z)&=&(2\overline z)^{-1} \HH^\pm_{1,1}(0,0)+\cdots\label{Hpm}\\
\VVV^\pm_{\frac12,1}(z,\overline z)\VVV^\mp_{\frac12,1}(-z,-\overline z)&=&\pm|2\overline z|^{-2} \HH_{0,1}(0,0)+\cdots.\label{H0}
\ee
In deriving the above expression we have also assumed that the OPE  $\{\VVV_{\frac12,1}\,\VVV_{1,1}\}$ is regular in the holomorphic side. This is generically true when
the $\N=2$ decomposition is available (as we are assuming here), see \cite{Vosmera:2019mzw} or appendix \ref{app:C} for a proof. Notice in particular that if no field is found at those particular singularities in the OPE, then the quartic effective potential is identically vanishing. In the next section we will encounter explicit examples of vanishing and non-vanishing auxiliary fields.

Given the above form of the localized fields, we can finally write the effective quartic potential (which coincides with the full effective potential up to quartic order) with simple matter 2-point functions of the auxiliary fields
\be
S_{\rm eff}(\varphi_A)=\frac14\Big(\langle\HH_{1,1}^+|\HH_{1,1}^-\rangle+\langle\HH_{0,1}|\HH_{0,1}\rangle\Big)+\mathcal{O}(\varphi_A^5).\label{SeffAux}
\ee
This expression is universal, only based on the assumption that $\varphi_A=\varphi^++\varphi^-$ and otherwise is independent on the ($\N=2$) closed string
background under consideration.

Notice also that \eqref{SeffAux}  is manifestly positive definite. Therefore flat directions of the potential are automatically  minima of the full effective action up to quartic order\footnote{Assuming that setting to zero the ghost dilaton $\varphi_D$ (which we have discarded) is a consistent truncation. This should follow from the fact that, by the dilaton theorem, 4-point couplings of $n$ $\varphi_D$'s and $(4-n)$ $\varphi_A$'s should be related to self-couplings of $(4-n)$  $\varphi_A$'s, which however vanish when $\N=2$ is present and $\varphi_A=\varphi^++\varphi^-$.}.

Then, just as in the open string analysis of \cite{Maccaferri:2018vwo, Maccaferri:2019ogq, Vosmera:2019mzw}, flat directions are controlled by the vanishing of the three ADHM-like constraints
\begin{align}
\HH_{1,1}^\pm&=0,\label{ADHM1}\\
\HH_{0,1}&=0.\label{ADHM2}
\end{align}

But,  by our construction in section \ref{sec221}, minima of the effective potential uplift to full SFT solutions and therefore the above constraints (\ref{ADHM1}, \ref{ADHM2}) appear to be  necessary and sufficient to guarantee exact marginality of the deformation triggered by $\varphi_A=\varphi^++\varphi^-$, up to quartic order.
% $and$ assuming that the ghost dilaton $\varphi_D$ decouples.
%%%%%%%%%%%%%%%%%%%%%%%%%%%%%%
\section{Example: Yang-Mills in flat space}\label{sec4}
%%%%%%%%%%%%%%%%%%%%%%%%%%%%%%
Let us now see what \eqref{SeffAux} reduces to in the simplest setting of the heterotic string in flat ten-dimensional space-time.

The massless zero momentum fields (except the universal ghost dilaton) can all be assembled in
\begin{align}
    \varphi_A = (g_{\mu\nu}+B_{\mu\nu})\xi c \psi^\mu e^{-\phi}\overline c\, i\overline{\p} \overline{X}^\nu+  A_{\mu i}\xi c \psi^\mu e^{-\phi}\overline{c}\overline{J}^i\,,
\end{align}
where $\mu=1,\ldots, 10$ runs over the spacetime dimensions and the adjoint gauge index $i=1,\ldots,\mathrm{dim}\,\mathfrak{g}$ runs over the currents $J^i$ of an affine Kac-Moody algebra $\widehat{\mathfrak{g}}_{k=1}$ which is specified by the given heterotic gauge group. The (hermitian) generators of the corresponding Lie algebra will be denoted by $T_i$ (where a representation has to be specified). They satisfy $[T_i,T_j]=if_{ijk}T_k$ and $\mathrm{tr}[T_iT_j]=2C\delta_{ij}$, where $f_{ijk}$ are the structure constants and $C$ is the Dynkin index of the representation at hand. We then have
\begin{align}
\overline J^i(\overline z)\overline J^j(\overline w)=\frac{\delta^{ij}}{(\overline z-\overline w)^2}+if_{ijk}\frac{\overline J^k(\overline w)}{\overline z-\overline w}+\cdots
\end{align}
%(for $\mathfrak{g}=\mathfrak{so}(32)$ we have $J^i=:\!\lambda^{r_i}\lambda^{s_i}\!:$, $r_i<s_i$ of the $\widehat{\mathfrak{so}}(32)_1$ AKM algebra (with $\lambda^r$ for $r=1,\ldots,32$ free fermions).
Let us introduce the complexified free fields
\begin{subequations}
\begin{align}
%	\p X^{a\pm} &= \frac{1}{\sqrt{2}}(\p X^{2a-1}\pm i \p X^{2a})\,,\\
    \psi^{a\pm} &= \frac{1}{\sqrt{2}}(\psi^{2a-1}\pm i \psi^{2a})\,,
\end{align}
\end{subequations}
where $a=1,\ldots,5$, so that
\begin{subequations}
\begin{align}
g_{\mu\nu}\psi^\mu  &= g_{(a+)\nu}\psi^{a+}+ g_{(a-)\nu}\psi^{a-}\,,\\
B_{\mu\nu}\psi^\mu  &= B_{(a+)\nu}\psi^{a+}+ B_{(a-)\nu}\psi^{a-}\,,\\
A_{\mu i}\psi^\mu &= A_{(a+)i}\psi^{a+}+ A_{(a-)i}\psi^{a-}
\end{align}
\end{subequations}
with
\begin{subequations}
\begin{align}
	g_{(a\pm)\nu} &= \frac{1}{\sqrt{2}}(g_{(2a-1)\nu}\mp i g_{(2a)\nu})\,,\\
	B_{(a\pm)\nu} &= \frac{1}{\sqrt{2}}(B_{(2a-1)\nu}\mp i B_{(2a)\nu})\,,\\
    A_{(a\pm)i}   &= \frac{1}{\sqrt{2}}(A_{(2a-1)i}\mp i A_{(2a)i})\,.
\end{align}
\end{subequations}
We clearly have the $\mathcal{N}=2$ decomposition
\begin{align}
\varphi_A=\varphi_A^++\varphi_A^-
\end{align}
with respect to the free-field $\mathcal{N}=2$ superconformal algebra with $R$-charge
\begin{align}
j = \sum_{a=1}^5 :\!\psi^{a-}\psi^{a+}\!:\,.
\end{align}
Explicitly we have
\begin{align}
\varphi_A^\pm &= (g_{(a\pm)\nu}+B_{(a\pm)\nu})\xi c \psi^{a\pm} e^{-\phi} i\overline{\p} \overline{X}^\nu+  A_{(a\pm) i}\xi c \psi^{a\pm} e^{-\phi}\overline{c}\overline{J}^i\,.
\end{align}
From \eqref{Hpm},\eqref{H0} we can compute the auxiliary fields
\begingroup\allowdisplaybreaks
\begin{subequations}
\label{eq:aux}
\begin{align}
    \mathbb{H}_{1,1}^\pm&=i\tensor{f}{^{ij}_k}A_{(a\pm)i}A_{(b\pm)j}\,:\!\psi^{a\pm}\psi^{b\pm}\!: \overline{J}^k\,,\\
  \mathbb{H}_{0,1}&=i\tensor{f}{^{ij}_k}A_{(a+)i}A_{(b-)j}\delta^{ab} \overline{J}^k\,.
\end{align}
\end{subequations}
\endgroup
Notice that $g_{\mu\nu}$ and $B_{\mu\nu}$ drop out  from the auxiliary fields because $\overline\p \overline X^\mu$ does not have a first-order pole in the OPE neither with itself nor with $\overline{J}^i$. A non-trivial auxiliary field is only obtained when the anti-holomorphic currents  entering in $\varphi_A$ are non-abelian!  From here we can already anticipate how the story will end, but it is nonetheless instructive to see how the expected  $\tr [A_\mu,A_\nu][A^\mu,A^\nu]$ potential is reconstructed, to appreciate the differences wrt the analogous open string  case in \cite{Maccaferri:2018vwo}, where the non-abelianity came from the Chan-Paton factors, rather than a non-abelian current algebra.

We substitute the expression \eqref{eq:aux} for the auxiliary fields into the general formula \eqref{SeffAux}. In order to ease the algebraic manipulations which are to follow, let us define the space-time matrix
\begin{align}
U =\text{diag}[u,u,u,u,u]\,,
\end{align}
where
\begin{align}
u = \frac{1}{\sqrt{2}}\begin{pmatrix}
1 & -i\\
1& + i
\end{pmatrix}\,,
\end{align}
together with the vectors
\begin{align}
\mathbf{A}_j&= (A_{1j},A_{2j},\ldots , A_{10j})\,,\\
\tilde{\mathbf{A}}_j &= (A_{(1+)j},A_{(1-)j},\ldots,A_{(5+)j}, A_{(5-)j})\,,
\end{align}
where $j$ is an adjoint index in the heterotic gauge algebra.
We can then write
\begin{align}
\tilde{\mathbf{A}}_j= U\mathbf{A}_j\,.
\end{align}
Also note that  the reality conditions on $A$ give that
\begin{align}
(\mathbf{A}_j)^\dagger= (\mathbf{A}_j)^T\,.
\end{align}
Let us also define
\begin{align}
V = \text{diag}[v,v,v,v,v]\,,
\end{align}
where
\begin{align}
v = \begin{pmatrix}
1 & 0\\
0& 0
\end{pmatrix}\,,
\end{align}
together with
\begin{align}
W = \text{diag}[w,w,w,w,w]\,,
\end{align}
where
\begin{align}
w = \begin{pmatrix}
 0 & -1\\
 1 & 0
\end{pmatrix}\,.
\end{align}
Using that $f_{ijk}=-f_{jik}$, we therefore obtain
\begingroup\allowdisplaybreaks
\begin{subequations}
\begin{align}
    S_\text{eff}^{(4)} &= \frac{1}{4}\tensor{f}{^{ij}_m}f^{klm}\bigg\{2A_{(a+)i}A_{(b+)j}A_{(a-)k}A_{(b-)l}-A_{(a+)i}A_{(a-)j}A_{(b+)k}A_{(b-)l} \bigg\}\,,\\
    &= \frac{1}{4}\tensor{f}{^{ij}_m}f^{klm}\bigg\{2\big[({\tilde{\mathbf{A}}}_k)^\dagger V\tilde{\mathbf{A}}_i\big]\big[(\tilde{\mathbf{A}}_l)^\dagger V\tilde{\mathbf{A}}_j\big] -\big[({\tilde{\mathbf{A}}}_j)^\dagger V\tilde{\mathbf{A}}_i\big]\big[(\tilde{\mathbf{A}}_l)^\dagger V\tilde{\mathbf{A}}_k\big] \bigg\}\,,\\
        &= \frac{1}{4}\tensor{f}{^{ij}_m}f^{klm}\bigg\{2\big[({{\mathbf{A}}}_k)^\dagger U^\dagger VU{\mathbf{A}}_i\big]\big[({\mathbf{A}}_l)^\dagger U^\dagger VU{\mathbf{A}}_j\big]+\nonumber\\
        &\hspace{4cm} -\big[({{\mathbf{A}}}_j)^\dagger U^\dagger VU{\mathbf{A}}_i\big]\big[({\mathbf{A}}_l)^\dagger U^\dagger VU {\mathbf{A}}_k\big] \bigg\}\,,\\
%        &= \frac{1}{2}\tensor{f}{^{ij}_m}f^{klm}\big[({{\mathbf{A}}}_k)^\dagger U^\dagger VU{\mathbf{A}}_i\big]\big[({\mathbf{A}}_l)^\dagger U^\dagger VU{\mathbf{A}}_j\big]+\nonumber\\
%        &\hspace{4cm} - \frac{1}{4}\tensor{f}{^{ij}_m}f^{klm}\big[({{\mathbf{A}}}_j)^\dagger U^\dagger VU{\mathbf{A}}_i\big]\big[({\mathbf{A}}_l)^\dagger U^\dagger VU {\mathbf{A}}_k\big] \,,\\
          &= +\frac{1}{8}\tensor{f}{^{ij}_m}f^{klm}\big[({{\mathbf{A}}}_k)^\dagger (I+iW){\mathbf{A}}_i\big]\big[({\mathbf{A}}_l)^\dagger (I+iW){\mathbf{A}}_j\big]+\nonumber\\
        &\hspace{4cm} - \frac{1}{16}\tensor{f}{^{ij}_m}f^{klm}\big[({{\mathbf{A}}}_j)^\dagger (I+iW){\mathbf{A}}_i\big]\big[({\mathbf{A}}_l)^\dagger (I+iW) {\mathbf{A}}_k\big] \,.
        \end{align}
        \end{subequations}
We can now use that fact that $(\mathbf{A}_j)^\dagger \mathbf{A}_i$ is symmetric in $i,j$ while $(\mathbf{A}_j)^\dagger W\mathbf{A}_i$ is antisymmetric in $i,j$ to rewrite
%\begin{align}
%S_\text{eff}^{(4)}&=     +\frac{1}{8}\tensor{f}{^{ij}_m}f^{klm}\big[(\mathbf{A}_k)^\dagger  \mathbf{A}_i \big]\big[(\mathbf{A}_l)^\dagger \mathbf{A}_j\big]+\nonumber\\
%          &\hspace{1cm}+\frac{i}{8}\tensor{f}{^{ij}_m}f^{klm}\Big\{ \big[(\mathbf{A}_k)^\dagger W \mathbf{A}_i \big]\big[(\mathbf{A}_l)^\dagger \mathbf{A}_j\big]+\big[(\mathbf{A}_k)^\dagger  \mathbf{A}_i\big]\big[ (\mathbf{A}_l)^\dagger W \mathbf{A}_j\big]\Big\}+\nonumber\\
%        &\hspace{1cm}+\frac{1}{8}\tensor{f}{^{ij}_m}f^{klm}\big[(\mathbf{A}_k)^\dagger W \mathbf{A}_i \big]\big[(\mathbf{A}_l)^\dagger W\mathbf{A}_j ]\big]+\nonumber\\
%        &\hspace{6cm}+ \frac{1}{16}\tensor{f}{^{ij}_m}f^{klm}\big[({{\mathbf{A}}}_j)^\dagger W{\mathbf{A}}_i\big]\big[({\mathbf{A}}_l)^\dagger W {\mathbf{A}}_k\big] \,.
%\end{align}
%Then, using again the symmetry and anti-symmetry of $(\mathbf{A}_j)^\dagger \mathbf{A}_i$ and $(\mathbf{A}_j)^\dagger W\mathbf{A}_i$, we can (anti)symmetrize the structure constants, thus obtaining
 %       \begin{subequations}
        \begin{align}
     S_\text{eff}^{(4)}
%   &=
%          +\frac{1}{8}\tensor{f}{^{ij}_m}f^{klm}\big[(\mathbf{A}_k)^\dagger  \mathbf{A}_i \big]\big[(\mathbf{A}_l)^\dagger \mathbf{A}_j\big]+\nonumber\\
%          &\hspace{1cm}+\frac{i}{16}(
%          \tensor{f}{^{ij}_m}f^{klm}+\tensor{f}{^{il}_m}f^{kjm}+\nonumber\\
%          &\hspace{4cm}-\tensor{f}{^{kj}_m}f^{ilm}-\tensor{f}{^{kl}_m}f^{ijm}
%          ) \big[(\mathbf{A}_k)^\dagger W \mathbf{A}_i \big]\big[(\mathbf{A}_l)^\dagger \mathbf{A}_j\big]+\nonumber\\
%        &\hspace{1cm}+\frac{1}{16}(2\tensor{f}{^{ij}_m}f^{klm}+\tensor{f}{^{ik}_m}f^{jlm}-2\tensor{f}{^{kj}_m}f^{ilm}-\tensor{f}{^{ki}_m}f^{jlm}+\nonumber\\
%        &\hspace{3cm}-2\tensor{f}{^{il}_m}f^{kjm}-\tensor{f}{^{ik}_m}f^{ljm}+2\tensor{f}{^{kl}_m}f^{ijm}+\tensor{f}{^{ki}_m}f^{ljm})\times\nonumber\\
%        &\hspace{5cm}\times\big[(\mathbf{A}_k)^\dagger W \mathbf{A}_i \big]\big[(\mathbf{A}_l)^\dagger W\mathbf{A}_j ]\big]\\
          &=
          +\frac{1}{8}\tensor{f}{^{ij}_m}f^{klm}\big[(\mathbf{A}_k)^\dagger  \mathbf{A}_i \big]\big[(\mathbf{A}_l)^\dagger \mathbf{A}_j\big]+\nonumber\\
        &\hspace{1cm}+\frac{1}{4}(\tensor{f}{^{ij}_m}f^{klm}+\tensor{f}{^{ik}_m}f^{jlm}+\tensor{f}{^{il}_{m}}\tensor{f}{^{jkm}})\big[(\mathbf{A}_k)^\dagger W \mathbf{A}_i \big]\big[(\mathbf{A}_l)^\dagger W\mathbf{A}_j ]\big]\,,
        \end{align}
%        \end{subequations}
where we note that
\begin{align}
\tensor{f}{^{ij}_m}f^{klm}+\tensor{f}{^{ik}_m}f^{jlm}+\tensor{f}{^{il}_{m}}\tensor{f}{^{jkm}}=0
\end{align}
by the Jacobi identity. We therefore end up with
\begin{align}
S_\text{eff}^{(4)}    &=+\frac{1}{8}\tensor{f}{^{ij}_m}A_{\mu i}A_{\nu j}\,\tensor{f}{_{kl}^{m}}A^{\mu k}A^{\nu l}\,.
\end{align}
This can be further rewritten as
%\begin{subequations}
\begin{align}
S_\text{eff}^{(4)} &= -\frac{1}{16C}\mathrm{tr}\big[[A_\mu,A_\nu][A^\mu, A^\nu]\big]\,,
\end{align}
%\end{subequations}
where in the last line we have denoted $A_\mu = A_{\mu i}T^i$. We have thus recovered the quartic potential of the heterotic gauge fields.
Notice that our (tree-level) construction did not depend on the details of the heterotic gauge group -- these are determined by modular invariance at one loop.
%\textcolor{red}{Moreover, note that this term was found in \cite{Gross:1986mw} by requiring gauge covariance of the kinetic term and $3$-point interaction and not computed directly.}

%where $T^i$ are the generators of $\mathfrak{g}$ (so that $[T_i,T_j]=if_{ijk}T^k$) in some representation with Dynkin index $C_\mathfrak{g}$ (so that $\text{tr}[T_mT_n] =2C_\mathfrak{g}\delta_{mn}$). Notice that as anticipated, our tree-level computation is insensitive to the particular nature of $\mathfrak{g}$ which has to be fixed by requiring consistency (modular invariance) at one loop.

%%%%%%%%%%%%%%%%%%%%%%%%%
\section{Conclusions}\label{concl}\label{sec5}
%%%%%%%%%%%%%%%%%%%%%%%%%%
In this paper we have developed  new computational tools to extract algebraic couplings in the tree-level effective action of the heterotic string from closed string field theory. We have
found that the same localization mechanism present in the open superstring \cite{Maccaferri:2018vwo, Maccaferri:2019ogq, Vosmera:2019mzw} is also at work here. This may appear rather surprising because we would in general expect  closed string physics to be  rather different from open string one.
However it is certainly less surprising from the  point of view of string field theory. In particular, especially in the small Hilbert space,  it turns out that open, closed and open-closed string field theories have all essentially the same algebraic structure encoded into a cyclic $L_\infty$ algebra (which is realized differently depending on the model under study).  Therefore also the derived effective actions should clearly have common universal features. We will elaborate more on this in \cite{Erbin:2019xx}.
In the particular case of the heterotic SFT, we expect that the localized form of the zero-momentum quartic effective action obtained in this paper using the WZW-like large Hilbert space formulation should agree with the corresponding result one would obtain within the $L_\infty$-based small Hilbert space formulation \cite{Erler:2014eba} (in the same way as it worked for the open superstring in \cite{Maccaferri:2019ogq}). Indeed, one expects both actions to be completely equivalent at the classical level, as it has already been shown for the open superstring \cite{Erler:2015uba, Erler:2015uoa, Erler:2017onq  } (see in particular the partially gauge fixed WZW-like action \eqref{eq:wzw-partial-fixed}).

A quite important conceptual point emerging from our work is that the massless equation of motion derived from the effective action are sufficient for establishing the existence of a full string field theory solution. This is a quite convenient reduced set of equations that can provide a useful alternative approach to exact marginality  and perhaps RG-flows triggered by marginally relevant operators (for both bulk and boundary degrees of freedom) where the standard worldsheet approaches (for example \cite{Recknagel:1998ih}) are not of much help.

In the  future it would be interesting to extend the localization method to Type-II theories and finally to the full open-closed theories, where both open and closed string moduli can genuinely  fluctuate. A first step into the interplay of open and closed strings consists in adding Ellwood Invariants \cite{Ellwood:2008jh} to the open string action and derive effective open-closed couplings by integrating out the massive open strings, slightly generalizing \cite{Maccaferri:2018vwo}. We plan to report on this soon \cite{MV}.  Another issue we have not touched until now is whether the localization mechanism can be useful for couplings involving space-time fermions. Needless to say, it would  be also interesting to explore the  localization of couplings beyond the quartic order and at finite momentum which is relevant for derivative couplings. The localization at the boundary of worldsheet moduli space and the needed $\N=2$ structure is naturally calling for a relation with topological strings \cite{holom, BCOV, Antoniadis:1993ze} which would be quite interesting to explore.

On a more general direction, it would be also interesting to explore the  possible existence of new type of localization mechanisms, not necessarily related to a
worldsheet $\N=2$. Some of them may involve non-trivially various currents in the (super) ghost sector and this may be useful to address the exact computation of the ghost-dilaton couplings which we left essentially untouched in this paper.

We hope our  progress will be useful in developing  ways to efficiently extract non trivial physical information from SFT, when the standard world-sheet methods fall short.

%%%%%%%%%%%%%%%%%%%%%%%%%%%%
\section*{Acknowledgments}
%%%%%%%%%%%%%%%%%%%%%%%%%%%%
We thank  Alberto Merlano for discussions and early collaboration on some of the fundamental ideas of this work. We thank Ted Erler and Martin Schnabl for useful discussions and comments.
We thank the Galileo Galilei Institute for having hosted the workshop ``String Theory from a worldsheet perspective" spring 2019,  which provided an invaluable
environment where some of these ideas have been discussed. CM thanks the Czech Academy of Science for hospitality in the initial stages of this work. JV also thanks INFN Turin for their hospitality during the later stages of this work.
The research of JV has been supported by the Czech Science Foundation - GA\v{C}R, project 19-06342Y and also by ESIF and M\v{S}MT (Project CoGraDS -CZ.02.1.01/0.0/0.0/15\_ 003/0000437). The work of HE and CM is partially supported by the MIUR PRIN
Contract 2015MP2CX4 ``Non-perturbative Aspects Of Gauge Theories And Strings''.

\appendix

%%%%%%%%%%%%%%
\section{Analysis of the massive out-of-gauge constraints}\label{app:A}
%%%%%%%%%%%%%%
Here we show that the two gauge constraints \eqref{eq:EOMRtH} and \eqref{eq:EOMRtQ} are automatically satisfied provided that we substitute the solution $R(\varphi)$ of $\text{EOM}_R$ and we also assume that the massless equation of motion $\text{EOM}_\varphi$ is solved. So in the following we will assume that
\begin{align}
\text{EOM}_\varphi(\varphi,R(\varphi))=\text{EOM}_R(\varphi,R(\varphi))=0\,.
\end{align}
We will also find it convenient to denote the sum of the two gauge constraints by
\begin{align}
\mathcal{GC}(\varphi,R)=\text{EOM}_{\tilde{\mathcal{R}}}(\varphi,R)+\text{EOM}_{\tilde{{R}}}(\varphi,R)\,.
\end{align}
Starting with $\text{EOM}_{\tilde{\mathcal{R}}}$, we first note that we can bring the $\eta_0$ from the projector $\Pi_\eta$ inside the interaction part to obtain
\begin{align}
\text{EOM}_{\tilde{\mathcal{R}}}(\varphi,R) &= -\overline{P}_0\xi_0\bigg\{\frac{1}{2!} [\eta_0 \Phi,\eta_0 Q\Phi]+\frac{1}{3!}\bigg(
  2 [\eta_0 \Phi,\eta_0 Q\Phi,Q\Phi]
	-\frac{3}{2}[\eta_0 \Phi,[\eta_0\Phi,Q\Phi]]+\nonumber\\
	&\hspace{1.8cm}  +\frac{1}{2}[\eta_0\Phi,[\Phi,\eta_0 Q\Phi]] -[\Phi,[\eta_0\Phi,\eta_0 Q\Phi]]
\bigg)+\mathcal{O}(\Phi^4)\bigg\}\bigg|_{\Phi=\varphi+R}\,.\label{eq:EOMRtEta1}
\end{align}
Now, using the definition of $\text{EOM}(\Phi)$, we can rewrite
\begin{subequations}
\begin{align}
\eta_0 Q\Phi \big|_{\Phi=\varphi+R(\varphi)}
&= \mathcal{J}(\Phi)\big|_{\Phi=\varphi+R(\varphi)}-\text{EOM}(\Phi)\big|_{\Phi=\varphi+R(\varphi)}\\
&=\mathcal{J}(\Phi)\big|_{\Phi=\varphi+R(\varphi)}-\mathcal{GC}(\varphi,R(\varphi))\,,\label{eq:etaQPhi}
\end{align}
\end{subequations}
where to go to the second line, we have used that $\text{EOM}_R(\varphi,R(\varphi))=\text{EOM}_\varphi(\varphi,R(\varphi))=0$, as per our assumptions. Substituting \eqref{eq:etaQPhi} into \eqref{eq:EOMRtEta1} and keeping only terms up to cubic order in $\Phi$, we obtain
\begin{align}
\text{EOM}_{\tilde{\mathcal{R}}}(\varphi,R(\varphi))&=
\overline{P}_0\xi_0\bigg\{\frac{1}{2!} [\eta_0 \Phi,\mathcal{GC}(\varphi,R(\varphi))]+\nonumber\\
	&\hspace{2cm} +\frac{1}{3!}\bigg(
  2 [\eta_0 \Phi,\mathcal{GC}(\varphi,R(\varphi)),Q\Phi]
	+\nonumber\\
	&\hspace{4cm} +\frac{1}{2}[\eta_0\Phi,[\Phi,\mathcal{GC}(\varphi,R(\varphi))]] +\nonumber\\
	 &\hspace{4.0cm}-[\Phi,[\eta_0\Phi,\mathcal{GC}(\varphi,R(\varphi))]]
\bigg)\bigg\}\bigg|_{\Phi=\varphi+R(\varphi)}
+\nonumber\\
&\hspace{1cm}-\overline{P}_0\xi_0\frac{1}{2!} [\eta_0 \Phi,\mathcal{J}(\Phi)]\big|_{\Phi=\varphi+R(\varphi)}+\nonumber\\
&\hspace{2cm}+\frac{1}{3!}\frac{3}{2}\overline{P}_0\xi_0[\eta_0 \Phi,[\eta_0\Phi,Q\Phi]]\big|_{\Phi=\varphi+R(\varphi)}+\mathcal{O}(\Phi^4)\,,
\end{align}
where we note that the quadratic part of $\mathcal{J}(\Phi)$ contributes to cancel the last term, namely
\begin{align}
-\overline{P}_0\xi_0\frac{1}{2} [\eta_0 \Phi,\mathcal{J}(\Phi)]+\frac{1}{4}\overline{P}_0\xi_0[\eta_0 \Phi,[\eta_0\Phi,Q\Phi]]=\mathcal{O}(\Phi^4)\,.
\end{align}
That is, defining the linear operator
\begin{align}
\mathcal{F}_1[A] &= \overline{P}_0\xi_0\bigg\{\frac{1}{2!} [\eta_0 \Phi,A] +\frac{1}{3!}\bigg(
  2 [\eta_0 \Phi,A,Q\Phi]
	 +\frac{1}{2}[\eta_0\Phi,[\Phi,A]] +\nonumber\\&\hspace{6cm}-[\Phi,[\eta_0\Phi,A]]
\bigg)+\mathcal{O}(\Phi^4)\bigg\}\bigg|_{\Phi=\varphi+R(\varphi)}\,,
\end{align}
we obtain the equation
\begin{align}
\text{EOM}_{\tilde{\mathcal{R}}}(\varphi,R(\varphi))=\mathcal{F}_1\big[\text{EOM}_{\tilde{\mathcal{R}}}(\varphi,R(\varphi))+\text{EOM}_{\tilde{{R}}}(\varphi,R(\varphi))\big]\,.
\end{align}
Note that using linearity of $\mathcal{F}_1$ we may rewrite this as
\begin{align}
(1-\mathcal{F}_1)\big[\text{EOM}_{\tilde{\mathcal{R}}}(\varphi,R(\varphi))\big]-\mathcal{F}_1\big[\text{EOM}_{\tilde{{R}}}(\varphi,R(\varphi))\big]\big]=0\,.
\label{eq:comp1}
\end{align}
Second, let us focus on the second gauge constraint, namely the out-of-Siegel equation
\begin{align}
\text{EOM}_{\tilde{R}}(\varphi,R(\varphi))
&=-b_0^+ c_0^+\eta_0\xi_0\eta_0 QR(\varphi)+\overline{P}_0 b_0^+c_0^+\eta_0\xi_0\mathcal{J}(\Phi)\big|_{\Phi=\varphi+R(\varphi)}\,.
\label{eq:EOMRtQA}
\end{align}
First, note that we have
\begin{align}
\eta_0 Q R(\varphi) &= \overline{P}_0 \mathcal{J}(\Phi)_{\Phi=\varphi+R(\varphi)} -\frac{b_0^+}{L_0^+}\overline{P}_0 Q\mathcal{J}(\Phi)_{\Phi=\varphi+R(\varphi)}+\nonumber\\
&\hspace{6cm}-Q\frac{b_0^+}{L_0^+}\overline{P}_0\xi_0 \eta_0 \mathcal{J}(\Phi)_{\Phi=\varphi+R(\varphi)}
\label{eq:etaQRphi}
\end{align}
so that after substituting \eqref{eq:etaQRphi} into \eqref{eq:EOMRtQA}, the gauge constraint may be rewritten as
\begin{align}
\text{EOM}_{\tilde{R}}(\varphi,R(\varphi)) &= -\eta_0\xi_0\frac{b_0^+}{L_0^+}\overline{P}_0 Q\mathcal{J}(\Phi)_{\Phi=\varphi+R(\varphi)}+\nonumber\\
&\hspace{4cm}-b_0^+ c_0^+ \eta_0\xi_0 Q\frac{b_0^+}{L_0^+}\overline{P}_0\xi_0 \eta_0 \mathcal{J}(\Phi)_{\Phi=\varphi+R(\varphi)} \,.\label{eq:EOMRtB}
\end{align}
Note that the second term has essentially already been dealt with above where, assuming that $\text{EOM}_\varphi(\varphi,R(\varphi))=0$, we have shown that
\begin{align}
\overline{P}\xi_0\eta_0 \mathcal{J}(\Phi)\big|_{\Phi=\varphi+R(\varphi)} = \mathcal{F}_1\big[\text{EOM}_{\tilde{\mathcal{R}}}(\varphi,R(\varphi))\big]+\mathcal{F}_1\big[\text{EOM}_{\tilde{{R}}}(\varphi,R(\varphi))\big]\,.
\end{align}
As for the first term in \eqref{eq:EOMRtB}, we first note that
\begingroup\allowdisplaybreaks
\begin{subequations}
\begin{align}
Q\mathcal{J}(\Phi) &= +\frac{1}{2!} Q[\eta_0 \Phi,Q\Phi]+\frac{1}{3!}\bigg(
   Q[\eta_0 \Phi,Q\Phi,Q\Phi]+\nonumber\\
   &\hspace{1cm}
   +Q[\eta_0 \Phi,[\Phi,Q\Phi]]
   -\frac{1}{2}Q[\Phi,[\Phi,\eta_0 Q\Phi]]
   -\frac{1}{2}Q[ \Phi,[Q\Phi,\eta_0 \Phi]]
\bigg)+\mathcal{O}(\Phi^4)\\
&= +\frac{1}{2!} [\eta_0 Q\Phi,Q\Phi]+\frac{1}{3!}\bigg(
[\eta_0 Q\Phi,Q\Phi,Q\Phi]
-[\eta_0 \Phi,[Q\Phi,Q\Phi]+\nonumber\\
&\hspace{9cm}
-2[Q\Phi,[Q\Phi,\eta_0 \Phi]]
   +\nonumber\\
   &\hspace{1cm}
   +[\eta_0 Q\Phi,[\Phi,Q\Phi]]
   +[\eta_0 \Phi,[Q\Phi,Q\Phi]]+\nonumber\\[+2mm]
   &\hspace{3cm}+\frac{1}{2}[Q\Phi,[\Phi,\eta_0 Q\Phi]]
   +\frac{1}{2}[\Phi,[Q\Phi,\eta_0 Q\Phi]]+\nonumber\\
   &\hspace{3cm}
   +\frac{1}{2}[Q \Phi,[Q\Phi,\eta_0 \Phi]]
   -\frac{1}{2}[ \Phi,[Q\Phi,\eta_0 Q\Phi]]
\bigg)+\mathcal{O}(\Phi^4)\\
&= +\frac{1}{2!} [\eta_0 Q\Phi,Q\Phi]+\frac{1}{3!}\bigg(
[\eta_0 Q\Phi,Q\Phi,Q\Phi]  +[\eta_0 Q\Phi,[\Phi,Q\Phi]]
   +\nonumber\\
   &\hspace{1cm}   -\frac{1}{2}[ \Phi,[Q\Phi,\eta_0 Q\Phi]]
%  +\nonumber\\[+2mm]
%   &\hspace{3cm}
   +\frac{1}{2}[Q\Phi,[\Phi,\eta_0 Q\Phi]]
   +\frac{1}{2}[\Phi,[Q\Phi,\eta_0 Q\Phi]]+\nonumber\\
   &\hspace{7.0cm}
-\frac{3}{2}[Q\Phi,[Q\Phi,\eta_0 \Phi]]
\bigg)+\mathcal{O}(\Phi^4)\,.
\end{align}
\end{subequations}
\endgroup
At this point, we can evaluate everything at $\Phi=\varphi+R(\varphi)$ and use the result \eqref{eq:etaQPhi} to find that (keeping only terms up to cubic order in $\Phi$)
\begin{align}
Q\mathcal{J}(\Phi)\big|_{\Phi=\varphi+R(\varphi)}&=-\bigg\{\frac{1}{2!} [\mathcal{GC}(\varphi,R(\varphi)),Q\Phi]+\frac{1}{3!}\bigg(
[\mathcal{GC}(\varphi,R(\varphi)),Q\Phi,Q\Phi] +\nonumber\\
&\hspace{6.5cm} +[\mathcal{GC}(\varphi,R(\varphi)),[\Phi,Q\Phi]]
   +\nonumber\\
   &\hspace{1cm}   -\frac{1}{2}[ \Phi,[Q\Phi,\mathcal{GC}(\varphi,R(\varphi))]]
%  +\nonumber\\[+2mm]
%   &\hspace{3cm}
   +\frac{1}{2}[Q\Phi,[\Phi,\mathcal{GC}(\varphi,R(\varphi))]]+\nonumber\\
   &\hspace{3.5cm}
   +\frac{1}{2}[\Phi,[Q\Phi,\mathcal{GC}(\varphi,R(\varphi))]]\bigg)\bigg\}\bigg|_{\Phi=\varphi+R(\varphi)}+\nonumber\\&\hspace{0.5cm}+\bigg\{\frac{1}{2!} [\mathcal{J}(\Phi),Q\Phi]-\frac{1}{3!}\frac{3}{2}[Q\Phi,[Q\Phi,\eta_0 \Phi]]\bigg\}\bigg|_{\Phi=\varphi+R(\varphi)}+\mathcal{O}(\Phi^4)\,.
\end{align}
where we note that the quadratic part of $\mathcal{J}(\Phi)$ will cancel the last term, that is
\begin{align}
\frac{1}{2} [\mathcal{J}(\Phi),Q\Phi]-\frac{1}{4}[Q\Phi,[Q\Phi,\eta_0 \Phi]]=\mathcal{O}(\Phi^4)\,.
\end{align}
Putting everything together, we observe that if we define the linear functional
\begingroup\allowdisplaybreaks
\begin{align}
\mathcal{F}_2[A] &=\Bigg\{\eta_0\xi_0\frac{b_0^+}{L_0^+}\overline{P}_0 \bigg[\frac{1}{2!} [A,Q\Phi]+\frac{1}{3!}\bigg(
[A,Q\Phi,Q\Phi] +[A,[\Phi,Q\Phi]]
   +\nonumber\\
   &\hspace{2.0cm}   -\frac{1}{2}[ \Phi,[Q\Phi,A]]
%  +\nonumber\\[+2mm]
%   &\hspace{3cm}
   +\frac{1}{2}[Q\Phi,[\Phi,A]]+ \frac{1}{2}[\Phi,[Q\Phi,A]]\bigg)\bigg]+\nonumber\\
   &\hspace{5.5cm}
-b_0^+ c_0^+ \eta_0\xi_0 Q\frac{b_0^+}{L_0^+}\mathcal{F}_1[A]+\mathcal{O}(\Phi^4)\Bigg\}\Bigg|_{\Phi=\varphi+R(\varphi)}
\end{align}
\endgroup
then we can substitute into \eqref{eq:EOMRtB} and write
\begin{align}
\text{EOM}_{\tilde{R}}(\varphi,R(\varphi)) = \mathcal{F}_2\big[\text{EOM}_{\tilde{\mathcal{R}}}(\varphi,R(\varphi))+\text{EOM}_{\tilde{R}}(\varphi,R(\varphi))\big]\,.
\end{align}
Combining this with our previous result \eqref{eq:comp1}, we therefore obtain the matrix equation
\begin{align}
\begin{pmatrix}
1-\mathcal{F}_1 & -\mathcal{F}_1\\
-\mathcal{F}_2 & 1-\mathcal{F}_2
\end{pmatrix}\begin{pmatrix}
\text{EOM}_{\tilde{\mathcal{R}}}(\varphi,R(\varphi))\\
\text{EOM}_{\tilde{R}}(\varphi,R(\varphi))
\end{pmatrix}=0\,.
\end{align}
Assuming that $\varphi$ is small, it is straightforward to expand the operator
\begin{align}
\mathfrak{F}=\begin{pmatrix}
1-\mathcal{F}_1 & -\mathcal{F}_1\\
-\mathcal{F}_2 & 1-\mathcal{F}_2
\end{pmatrix}
\end{align}
in powers of $\varphi$ and thereby show that $\mathfrak{F}$ is in fact perturbatively invertible. Hence, given that we are only interested in the perturbative expansion of the effective action in powers of $\varphi$, we can conclude that $\text{EOM}_\varphi(\varphi,R(\varphi))=0$ and $\text{EOM}_\mathcal{R}(\varphi,R(\varphi))=0$ imply that $\text{EOM}_{\tilde{\mathcal{R}}}(\varphi,R(\varphi))=0$ and $\text{EOM}_{\tilde{{R}}}(\varphi,R(\varphi))=0$.
\section{Reducing the kernel of $L_0^+$}\label{app:B}
%%%%%%%%%%%%%%%%%%%%%%%%%%%%%%%%%%
In this appendix we  discuss how to  reduce the field content of $\text{ker}\,L_0^+$ given by \eqref{kernel}, to the truly physical fields.
This will be a three-step process: first we will partially gauge fix the effective action \eqref{seff} by setting $\varphi_3$ (the massless fields in the small Hilbert space) to zero. Consequently we will show that the missing equation of motion will be contained in the remaining part of the action. Then we will get rid of the Nakanishi-Lautrup field $\varphi_2$ by integrating it out and showing that this operation will not correct the physical couplings of the effective action up to quartic order. Finally, we will show that the pure gauge field $\varphi_B$ decouples from both the cubic and the quartic couplings.

\subsection{Partially gauge-fixing the effective action}\label{parg}

Here we will show that in the effective action for $\text{ker}\,L_0$ fields we can partially gauge-fix $\xi_0 \varphi=0$ to obtain a new effective action for fields which satisfy $\varphi=\xi_0\psi$ for some $\psi$ such that $\eta_0\psi=0$. But before we do this, let us discuss partial gauge fixing already at the level of the full heterotic SFT action. Indeed, if we write
\begin{align}
\Phi = \hat{\Phi}+ \Phi_3\,,
\end{align}
where $\hat{\Phi}=\Pi_\eta \Phi$ and $\Phi_3 = \overline{\Pi}_\eta\Phi$ and we partially gauge-fix by setting $\Phi_3$, then the respective equations of motion read
\begin{subequations}
\begin{align}
\text{EOM}_{\hat{\Phi}}(\hat{\Phi}) &= -\eta_0 Q \hat{\Phi} + \eta_0\xi_0\mathcal{J}(\hat{\Phi})\,,\\
\text{EOM}_{\Phi_3}(\hat{\Phi}) &= +\xi_0 \eta_0\mathcal{J}(\hat{\Phi})\,.
\end{align}
\end{subequations}
However, it is easy to see that up to order $\mathcal{O}(\Phi^4)$ the gauge constraint \eqref{eq:pgc} is already implied by the equation of motion $\text{EOM}_{\hat{\Phi}}(\hat{\Phi})$. Indeed, we can first bring $\eta_0$ inside $\mathcal{J}$ and then
assuming that $\text{EOM}_{\hat{\Phi}}(\hat{\Phi})=0$ we can substitute for $\eta_0 Q\hat{\Phi}$ in terms of $\mathcal{J}(\hat{\Phi})$. Realizing that $\mathcal{J}(\hat{\Phi})=\mathcal{O}(\hat{\Phi}^2)$ and ignoring $\mathcal{O}(\hat{\Phi}^4)$ terms, we finally obtain
\begingroup\allowdisplaybreaks
\begin{subequations}
\begin{align}
\text{EOM}_{\Phi_3}(\hat{\Phi}) &= -\xi_0\bigg\{\frac{1}{(2!)^2} [\eta_0 \hat{\Phi},\eta_0\xi_0 [\eta_0 \hat{\Phi},Q\hat{\Phi}]]-\frac{1}{3!}\frac{3}{2}
[\eta_0\hat{\Phi},[\eta_0\hat{\Phi},Q\hat{\Phi}]]\bigg\}+\mathcal{O}(\hat{\Phi}^4)\\[+1mm]
&=\mathcal{O}(\hat{\Phi}^4)\,,
\end{align}
\end{subequations}
\endgroup
where to deal with the first term, we have substituted $\eta_0 \xi_0 = 1-\xi_0 \eta_0$, brought $\eta_0$ inside the 2-string product and substituted for $\eta_0 Q\hat{\Phi}$. Hence, we obtain that $\text{EOM}_{\hat{\Phi}}(\hat{\Phi})=0$ implies $\text{EOM}_{\Phi_3}(\hat{\Phi})=0$ (at least up to quartic order). Therefore, at least up to quartic order, by partial gauge fixing of the full heterotic SFT, we do not incur any non-trivial gauge constraints. Writing
$\Phi = \xi_0 \Psi$ (dropping the hat over the partially gauge-fixed fields), the partially gauge-fixed action can be rewritten as
\begin{align}
S_{S}(\Psi)&=\frac{1}{2!}\big\langle  \Psi, Q\Psi\big\rangle_S+\frac{1}{3!}\big\langle\Psi,[\xi_0\Psi,Q\xi_0\Psi]\big\rangle+\nonumber\\
&\hspace{2cm}+\frac{1}{4!}\bigg(\big\langle \Psi,[\xi_0\Psi,Q\xi_0\Psi,Q\xi_0\Psi]\big\rangle+\big\langle \Psi,[\xi_0\Psi,[\xi_0\Psi,Q\xi_0\Psi]]\big\rangle\bigg)\nonumber\\
&\hspace{4cm}+\frac{1}{8}\bigg\langle[\Psi,Q\xi_0\Psi] , \frac{b_0^+}{L_0^+}\xi_0\overline{P}_0[\Psi,Q\xi_0\Psi]\bigg\rangle+O(\Psi^5)
\label{eq:wzw-partial-fixed}
\end{align}
Adopting the idea of \cite{Maccaferri:2019ogq}, this action can be completely rewritten (at least to the quartic order) using the small Hilbert space BPZ product $\big\langle\ldots\big\rangle_S$ and the picture raising operator $X_0$. Indeed we have
\begingroup\allowdisplaybreaks
\begin{subequations}
\begin{align}
\big\langle\Psi,[\xi_0\Psi,Q\xi_0\Psi]\big\rangle &= \big\langle\Psi,(\xi_0\eta_0+\eta_0\xi_0)[\xi_0\Psi,Q\xi_0\Psi]\big\rangle\\
&= -\big\langle\Psi,\eta_0[\xi_0\Psi,Q\xi_0\Psi]\big\rangle_S\\
&= \big\langle\Psi,[\Psi,X_0\Psi]\big\rangle_S+\big\langle\Psi,[\xi_0\Psi,Q\Psi]-[\Psi,\xi_0 Q\Psi]\big\rangle_S
\end{align}
\end{subequations}
\endgroup
together with
\begingroup\allowdisplaybreaks
\begin{subequations}
\begin{align}
\big\langle \Psi,[\xi_0\Psi,Q\xi_0\Psi,Q\xi_0\Psi]\big\rangle&=\big\langle \Psi,(\xi_0\eta_0+\eta_0\xi_0)[\xi_0\Psi,Q\xi_0\Psi,Q\xi_0\Psi]\big\rangle\\
&=-\big\langle \Psi,\eta_0[\xi_0\Psi,Q\xi_0\Psi,Q\xi_0\Psi]\big\rangle_S\\
&=
\big\langle \Psi,[\Psi,X_0\Psi,X_0\Psi]\big\rangle_S
+\nonumber\\
&\hspace{2cm}+\big\langle \Psi,2[\xi_0\Psi,Q\Psi,X_0\Psi]-2[\Psi,\xi_0 Q\Psi,X_0\Psi]\big\rangle_S+\nonumber\\
&\hspace{2cm}
+\big\langle \Psi,[\Psi,\xi_0Q\Psi,\xi_0Q\Psi]-2[\xi_0\Psi,Q\Psi,\xi_0Q\Psi]\big\rangle_S
\end{align}
\end{subequations}
\endgroup
and
\begingroup\allowdisplaybreaks
\begin{subequations}
\begin{align}
\big\langle \Psi,[\xi_0\Psi,[\xi_0\Psi,Q\xi_0\Psi]]\big\rangle &= \big\langle \Psi,(\xi_0\eta_0+\eta_0\xi_0)[\xi_0\Psi,[\xi_0\Psi,Q\xi_0\Psi]]\big\rangle\\
&= -\big\langle \Psi,\eta_0[\xi_0\Psi,[\xi_0\Psi,Q\xi_0\Psi]]\big\rangle_S\\
&= \big\langle \Psi,[\Psi,[\xi_0\Psi,Q\xi_0\Psi]]+[\xi_0\Psi,[\Psi,Q\xi_0\Psi]]+[\xi_0\Psi,[\xi_0\Psi,Q\Psi]]\big\rangle_S\\
&=  \big\langle \Psi,[\Psi,[\xi_0\Psi,X_0\Psi]]+[\xi_0\Psi,[\Psi,X_0\Psi]]\big\rangle_S+\nonumber\\
&\hspace{2.5cm}+\big\langle \Psi,[\xi_0\Psi,[\xi_0\Psi,Q\Psi]]-[\Psi,[\xi_0\Psi,\xi_0 Q\Psi]]+\nonumber\\
&\hspace{6.5cm}-[\xi_0\Psi,[\Psi,\xi_0 Q\Psi]]\big\rangle_S
\end{align}
\end{subequations}
\endgroup
That is, we can rewrite the action as
\begin{align}
S_{S}(\Psi)&=S_{S}^{(2)}(\Psi)+S_{S}^{(3)}(\Psi)+S_{S}^{(4)}(\Psi)+\mathcal{O}(\Psi^5)\,,
\end{align}
where
\begingroup\allowdisplaybreaks
\begin{subequations}
\begin{align}
S_{S}^{(2)}(\Psi) &= \frac{1}{2!}\big\langle\Psi,Q\Psi\big\rangle_S\,,\\
S_{S}^{(3)}(\Psi) &=\frac{1}{3!}\Big(\big\langle\Psi,[\Psi,X_0\Psi]\big\rangle_S+\big\langle\Psi,[\xi_0\Psi,Q\Psi]-[\Psi,\xi_0 Q\Psi]\big\rangle_S\Big)\,,\\
S_{S}^{(4)}(\Psi) &=\frac{1}{4!}\Big(\big\langle \Psi,[\Psi,X_0\Psi,X_0\Psi]\big\rangle_S
+\nonumber\\
&\hspace{2cm}+\big\langle \Psi,2[\xi_0\Psi,Q\Psi,X_0\Psi]-2[\Psi,\xi_0 Q\Psi,X_0\Psi]\big\rangle_S+\nonumber\\
&\hspace{2cm}
+\big\langle \Psi,[\Psi,\xi_0Q\Psi,\xi_0Q\Psi]-2[\xi_0\Psi,Q\Psi,\xi_0Q\Psi]\big\rangle_S+\nonumber\\
&\hspace{0.4cm} +\big\langle \Psi,[\Psi,[\xi_0\Psi,X_0\Psi]]+[\xi_0\Psi,[\Psi,X_0\Psi]]\big\rangle_S+\nonumber\\
&\hspace{2.5cm}+\big\langle \Psi,[\xi_0\Psi,[\xi_0\Psi,Q\Psi]]-[\Psi,[\xi_0\Psi,\xi_0 Q\Psi]]+\nonumber\\
&\hspace{6.5cm}-[\xi_0\Psi,[\Psi,\xi_0 Q\Psi]]\big\rangle_S\Big)\,.
\end{align}
\end{subequations}
\endgroup
It would be interesting to compare this action with the $L_\infty$-based theory \cite{Erler:2014eba} and construct the field redefinition mapping the two theories on one another (see \cite{Erler:2015uba,Erler:2015uoa} for similar results relating the Berkovits' WZW-like open superstring field theory and the  $A_\infty$-based theory).

Let us now repeat this partial gauge-fixing procedure in the case of the effective action for $\text{ker}\,L_0$ fields $\varphi$. To this end, we decompose
\begin{align}
\varphi = \hat{\varphi}+\varphi_3\,,
\end{align}
where we have $\hat{\varphi}=\Pi_\eta \varphi$ and $\varphi_3=\overline{\Pi}_\eta\varphi$. Fixing the $\xi_0$-gauge by setting ${\varphi_3}=0$, we obtain equations of motion
\begin{subequations}
\begin{align}
\text{eom}_{\hat{\varphi}}(\hat{\varphi}) &= \overline{\Pi}_\eta\text{eom}(\hat{\varphi}) = -\eta_0 Q\hat{\varphi} +\eta_0\xi_0 P_0\mathcal{J}(\hat{\varphi}+R(\hat{\varphi}))\,,\\
\text{eom}_{\varphi_3}(\hat{\varphi}) &= {\Pi}_\eta\text{eom}(\hat{\varphi}) = +\xi_0\eta_0 P_0\mathcal{J}(\hat{\varphi}+R(\hat{\varphi}))\,.\label{eq:pgc}
\end{align}
\end{subequations}
Bringing $\eta_0$ inside $\mathcal{J}$ and assuming that $\text{eom}_{\hat{\varphi}}(\hat{\varphi})=0$, one can show that
\begingroup\allowdisplaybreaks
\begin{subequations}
\begin{align}
\text{eom}_{\varphi_3}(\hat{\varphi}) &= \xi_0\bigg\{-\frac{1}{(2!)^2} P_0[\eta_0 \hat{\varphi},\eta_0\xi_0 P_0[\eta_0 \hat{\varphi},Q\hat{\varphi}]]+\frac{1}{3!}\frac{3}{2}P_0
[\eta_0\hat{\varphi},[\eta_0\hat{\varphi},Q\hat{\varphi}]]+\nonumber\\
&\hspace{5.5cm}
-\frac{1}{4}P_0 [\eta_0\hat{\varphi}, \overline{P}_0[\eta_0\hat{\varphi},Q\hat{\varphi}]]\bigg\}+\mathcal{O}(\hat{\varphi}^4)\\
%&= \xi_0\bigg\{-\frac{1}{(2!)^2} P_0[\eta_0 \hat{\varphi}, P_0[\eta_0 \hat{\varphi},Q\hat{\varphi}]]-\frac{1}{(2!)^2} P_0[\eta_0 \hat{\varphi},\xi_0 P_0[\eta_0 \hat{\varphi},\eta_0 Q\hat{\varphi}]]]+\nonumber\\
%&\hspace{2cm}+\frac{1}{3!}\frac{3}{2}P_0
%[\eta_0\hat{\varphi},[\eta_0\hat{\varphi},Q\hat{\varphi}]
%-\frac{1}{4}P_0[\eta_0\hat{\varphi}, \overline{P}_0[\eta_0\hat{\varphi},Q\hat{\varphi}]]\bigg\}+\mathcal{O}(\hat{\varphi}^4)\\
%&= \xi_0\bigg\{-\frac{1}{(2!)^2} P_0[\eta_0 \hat{\varphi}, P_0[\eta_0 \hat{\varphi},Q\hat{\varphi}]]+\frac{1}{4}P_0[\eta_0\hat{\varphi},{P}_0[\eta_0\hat{\varphi},Q\hat{\varphi}]]+\nonumber\\
%&\hspace{2cm}+\frac{1}{3!}\frac{3}{2}P_0
%[\eta_0\hat{\varphi},[\eta_0\hat{\varphi},Q\hat{\varphi}]
%-\frac{1}{4}P_0[\eta_0\hat{\varphi},[\eta_0\hat{\varphi},Q\hat{\varphi}]]\bigg\}+\mathcal{O}(\hat{\varphi}^4)\\
&=\mathcal{O}(\hat{\varphi}^4)\,,
\end{align}
\end{subequations}
\endgroup
where we again have to substitute $\eta_0\xi_0 = 1-\xi_0\eta_0$ and bring $\eta_0$ inside the 2-string product to deal with the first term. Therefore, at least up to quartic order, fixing the partial gauge for the effective action does not produce additional gauge constraints. Eventually, writing $\hat{\varphi}=\xi_0\hat{\psi}$, we find that the partially gauge-fixed effective action can be rewritten manifestly in the small Hilbert space as
\begin{align}\label{seffSHS}
S_{\text{eff},S}(\hat{\psi})&=S_{\text{eff},S}^{(2)}(\hat{\psi})+S_{\text{eff},S}^{(3)}(\hat{\psi})+S_{\text{eff},S}^{(4)}(\hat{\psi})+\mathcal{O}(\hat{\psi}^5)\,,
\end{align}
where
\begingroup\allowdisplaybreaks
\begin{subequations}
\begin{align}
S_{\text{eff},S}^{(2)}(\hat{\psi}) &= \frac{1}{2!}\big\langle\hat{\psi},Q\hat{\psi}\big\rangle_S\,,\\
S_{\text{eff},S}^{(3)}(\hat{\psi}) &=\frac{1}{3!}\Big(\big\langle\hat{\psi},[\hat{\psi},X_0\hat{\psi}]\big\rangle_S+\big\langle\hat{\psi},[\xi_0\hat{\psi},Q\hat{\psi}]-[\hat{\psi},\xi_0 Q\hat{\psi}]\big\rangle_S\Big)\,,\\
S_{\text{eff},S}^{(4)}(\hat{\psi}) &=\frac{1}{4!}\Big(\big\langle \hat{\psi},[\hat{\psi},X_0\hat{\psi},X_0\hat{\psi}]\big\rangle_S
+\nonumber\\
&\hspace{2cm}+\big\langle \hat{\psi},2[\xi_0\hat{\psi},Q\hat{\psi},X_0\hat{\psi}]-2[\hat{\psi},\xi_0 Q\hat{\psi},X_0\hat{\psi}]\big\rangle_S+\nonumber\\
&\hspace{2cm}
+\big\langle \hat{\psi},[\hat{\psi},\xi_0Q\hat{\psi},\xi_0Q\hat{\psi}]-2[\xi_0\hat{\psi},Q\hat{\psi},\xi_0Q\hat{\psi}]\big\rangle_S+\nonumber\\
&\hspace{0.4cm} +\big\langle \hat{\psi},[\hat{\psi},[\xi_0\hat{\psi},X_0\hat{\psi}]]+[\xi_0\hat{\psi},[\hat{\psi},X_0\hat{\psi}]]\big\rangle_S+\nonumber\\
&\hspace{2.5cm}+\big\langle \hat{\psi},[\xi_0\hat{\psi},[\xi_0\hat{\psi},Q\hat{\psi}]]-[\hat{\psi},[\xi_0\hat{\psi},\xi_0 Q\hat{\psi}]]+\nonumber\\
&\hspace{6.5cm}-[\xi_0\hat{\psi},[\hat{\psi},\xi_0 Q\hat{\psi}]]\big\rangle_S\Big)+\nonumber\\
&\hspace{0.4cm}-\frac{1}{8}\Bigg(\bigg\langle[\hat{\psi},X_0\hat{\psi}] , \frac{b_0^+}{L_0^+}\overline{P}_0[\hat{\psi},X_0\hat{\psi}]\bigg\rangle_S\nonumber\\
&\hspace{2.4cm}-2\bigg\langle[\hat{\psi},X_0\hat{\psi} ], \frac{b_0^+}{L_0^+}\overline{P}_0\Big([\hat{\psi},\xi_0 Q\hat{\psi}]+\xi_0[\hat{\psi},Q\hat{\psi}]\Big)\bigg\rangle_S\nonumber\\
&\hspace{2.4cm}+\bigg\langle\hat{\psi},\bigg[\xi_0 Q\hat{\psi} , \frac{b_0^+}{L_0^+}\overline{P}_0[\hat{\psi},\xi_0 Q\hat{\psi}]\bigg]\bigg\rangle_S\nonumber\\
&\hspace{2.4cm}
+\bigg\langle\hat{\psi},\bigg[\xi_0Q\hat{\psi} , \frac{b_0^+}{L_0^+}\xi_0\overline{P}_0[\hat{\psi},Q\hat{\psi}]\bigg]\bigg\rangle_S\nonumber\\
&\hspace{2.4cm}-\bigg\langle\hat{\psi},\bigg[Q\hat{\psi} , \frac{b_0^+}{L_0^+}\xi_0\overline{P}_0[\hat{\psi},\xi_0Q\hat{\psi}]\bigg]\bigg\rangle_S\Bigg)
\end{align}
\end{subequations}
\endgroup
Again, it would be interesting to discuss the relation of the WZW effective action written in the small Hilbert space and the effective action derived from the $L_\infty$-based  theory. In particular, one should be able to check whether this relation is consistent with the $P_0$ projection of the field redefinition relating the two theories.

\subsection{Integrating out the Nakanishi-Lautrup field}

 Next, we would like to get rid of the Nakanishi-Lautrup field $\varphi_2$. Noting that it can be written as $\varphi_2 = c_0^+\chi$ for some $\chi$, we may simply fix Siegel gauge $b_0^+ \varphi=0$, thereby setting $\varphi_2=0$. However, if we do this, then we end up with equations of motion
\begin{subequations}
\begin{align}
\text{eom}_{\varphi_1}(\varphi_1) &= -c_0^+ b_0^+\eta_0 Q\varphi_1 +c_0^+ b_0^+\eta_0  \xi_0P_0\mathcal{J}(\varphi_1+R(\varphi_1))\nonumber\\
&= c_0^+ b_0^+\eta_0  \xi_0P_0\mathcal{J}(\varphi_1+R(\varphi_1))\label{is}\\
\text{eom}_{\varphi_2}(\varphi_1) &= - b_0^+c_0^+\eta_0 Q\varphi_1 + b_0^+c_0^+\eta_0  \xi_0P_0\mathcal{J}(\varphi_1+R(\varphi_1))\nonumber\\
&= b_0^+ c_0^+ \eta_0  \xi_0P_0\mathcal{J}(\varphi_1+R(\varphi_1))\label{gc}
\end{align}
\end{subequations}
where we have used the fact that $\eta_0 Q\varphi_1=0$. Notice that the gauge constraint (out-of-Siegel equation) \eqref{gc} is given by a projection orthogonal to the one determining the in-Siegel equations of motion \eqref{is}. Therefore it does not seem that we can show that the out-of-Siegel equation is automatically solved assuming the in-Siegel equations. In this case the gauge-fixed action alone would not give complete information about the dynamics and it would be neccessary to supplement it by a non-trivial constraint. To overcome this difficulty we will therefore adopt a different approach: instead of setting $\varphi_2$  to zero, we will integrate it out. That is, we first solve the equation of motion
\begin{subequations}
\begin{align}
\text{eom}_{\varphi_2}(\varphi_1,\varphi_2) &= - b_0^+c_0^+\eta_0 b_0^+\mathcal{M}^+\varphi_2 + b_0^+c_0^+\eta_0  \xi_0P_0\mathcal{J}(\varphi_1+\varphi_2+R(\varphi_1+\varphi_2))\\
&=  b_0^+\eta_0\mathcal{M}^+\varphi_2 - b_0^+ \eta_0 c_0^+ \xi_0P_0\mathcal{J}(\varphi_1+\varphi_2+R(\varphi_1+\varphi_2)),
\end{align}
\end{subequations}
for $\varphi_2$ as a function of $\varphi_1$. In particular, we have
\begin{align}
\mathcal{M}^+ \varphi_2(\varphi_1) = c_0^+ \xi_0 P_0\mathcal{J}\big(\varphi_1+\varphi_2(\varphi_1)+R(\varphi_1+\varphi_2(\varphi_1))\big)\,.
\end{align}
At this point, we note that $\mathcal{M}^+$ together with operators $\mathcal{M}^- = M^-+\overline{M}^-$ and $\mathcal{M}_z=M_z+\overline{M}_z$, given by (see e.g.\ \cite{Asano:2016rxi})
\begin{subequations}
\begin{align}
M^- &= -\sum_{m=1}^\infty\frac{1}{2m}b_{-m}b_m+\sum_{q+\frac{1}{2}\in\mathbb{Z}_{>0}}\frac{1}{2}\beta_{-q}\beta_{q}\\
\overline{M}^-  &= -\sum_{m=1}^\infty\frac{1}{2m}\overline{b}_{-m}\overline{b}_m
\end{align}
\end{subequations}
and
\begin{subequations}
\begin{align}
M_z&=\frac{1}{2}\sum_{m\in\mathbb{Z}_{>0}}(c_{-m}b_m-b_{-m}c_m)-\frac{1}{2}\sum_{q+\frac{1}{2}\in\mathbb{Z}_{>0}}(\gamma_{-q}\beta_{q}+\beta_{-q}\gamma_{q})\,,\\
M_z&=\frac{1}{2}\sum_{m\in\mathbb{Z}_{>0}}(\overline{c}_{-m}\overline{b}_m-\overline{b}_{-m}\overline{c}_m)
\end{align}
\end{subequations}
form an $SU(1,1)$ algebra
\begin{subequations}
\begin{align}
[\mathcal{M}^+,\mathcal{M}^-] &=2\mathcal{M}_z\,,\\
[\mathcal{M}_z,\mathcal{M}^+] &= +\mathcal{M}^+\,,\\
[\mathcal{M}_z,\mathcal{M}^-] &= -\mathcal{M}^-\,.
\end{align}
\end{subequations}
Let us now show that we have $$\mathcal{M}^-\varphi_2=0.$$ To do this it is convenient to write
\begin{align}
\xi \p \xi e^{-2\phi} &= \gamma^{-2}\,,\\
\xi e^{-\phi} &= \gamma^{-1}
\end{align}
and then notice that
\begingroup\allowdisplaybreaks
\begin{subequations}
\begin{align}
\gamma^{-1}(0)\big|0\big\rangle  &= \Big(\gamma_{\frac{1}{2}}+\sum_{r\neq \frac{1}{2}}\gamma_r z^{-r+\frac{1}{2}}\Big)^{-1}\Big|_{z=0}\big| 0\big\rangle\\
&= (\gamma_{\frac{1}{2}})^{-1}\Big(1+(\gamma_{\frac{1}{2}})^{-1}\sum_{r\neq \frac{1}{2}}\gamma_r z^{-r+\frac{1}{2}}\Big)^{-1}\Big|_{z=0}\big| 0\big\rangle\\
&= (\gamma_{\frac{1}{2}})^{-1}\big|0 \big\rangle +(\gamma_{\frac{1}{2}})^{-1}\sum_{k=1}^\infty(-1)^k\Big((\gamma_{\frac{1}{2}})^{-1}\sum_{r\neq \frac{1}{2}}\gamma_r z^{-r+\frac{1}{2}}\Big)^{k}\Big|_{z=0}\big| 0\big\rangle\,,\\
&=(\gamma_{\frac{1}{2}})^{-1}\big|0 \big\rangle,
\end{align}
\end{subequations}
\endgroup
where to perform the last step we have noted that the oscillators in the sum $\sum_{r\neq 1/2}$ either kill $|0\rangle$ or they are multiplied by a positive power of $z$ so they vanish as $z\to 0$. Analogously we can show that
\begin{align}
\gamma^{-2}(0)|0\rangle = (\gamma_\frac{1}{2})^{-2}\big|0\big\rangle\,.
\end{align}
Therefore $\varphi_2$ is given by
\begin{align}
|\varphi_2\rangle = 2c_0^+\Big(c_1(\gamma_\frac{1}{2})^{-1}\big|\mathbf{V}_\frac{1}{2}\big\rangle + c_1\overline{c}_1(\gamma_\frac{1}{2})^{-2}\big|\overline{\mathbf{V}}_1\big\rangle\Big)\,,
\end{align}
from where it is easy to see that we have $\mathcal{M}^-|\varphi_2\rangle=0$.
We can therefore write
\begin{align}
\mathcal{M}^{-}\mathcal{M}^{+}\varphi_2=[\mathcal{M}^{-},\mathcal{M}^{+}]\varphi_2 = -2\mathcal{M}_z\varphi_2= \varphi_2\,,
\end{align}
where we have used that $\mathcal{M}_z = \frac{1}{2}\hat{N}_g$ where $\hat{N}_g$ measures the ghost number due to non-zero oscillator modes.\footnote{That is, the total ghost number $N_g$ can be written as
\begin{align}
N_g = \hat{N}_g+c_0 b_0 +\overline{c}_0\overline{b}_0+1\,.
\end{align}
}
We therefore obtain
\begin{align}
\varphi_2(\varphi_1) = \mathcal{M}^- c_0^+ \xi_0 P_0 \mathcal{J}\big(\varphi_1+\varphi_2(\varphi_1)+R(\varphi_1+\varphi_2(\varphi_1))\big)\,.
\end{align}
This is a recurrence relation for the function $\varphi_2(\varphi_1)$ and can be solved in the usual manner
\begin{align}
\varphi_2(\varphi_1) =  \frac{1}{2!}\mathcal{M}^- c_0^+ \xi_0 P_0 [\eta\varphi_1,Q\varphi_1]+\mathcal{O}(\varphi_1^3)\,.
\end{align}
However, we note that we can write
\begin{align}
\varphi_1 = \varepsilon_{ik}\xi c \mathbb{V}^i_\frac{1}{2}e^{-\phi} \overline{c}\overline{\mathbb{W}}^k_1+2(D+B)\xi c\eta+(D-B)c\xi\p \xi e^{-2\phi}\overline{c}\overline{\p}^2 \overline{c}  \,,
\end{align}
so that
\begin{subequations}
\begin{align}
\eta_0\varphi_1
%&= c \mathbb{V}_\frac{1}{2}e^{-\phi} \overline{c}\overline{\mathbb{V}}_1 +DY_0 Q(\p c-\overline{\p}\overline{c})+B Y_0 Q(\p c+\overline{\p}\overline{c})\nonumber\\
&= \varepsilon_{ik}c \mathbb{V}^i_\frac{1}{2}e^{-\phi} \overline{c}\overline{\mathbb{W}}^k_1+2(D+B)c\eta-(D-B)c \p \xi e^{-2\phi}\overline{c}\overline{\p}^2 \overline{c}\,,\\
Q\varphi_1
%&=  (c\mathbb{V}_1-e^{\phi}\eta\mathbb{V}_\frac{1}{2})\overline{c}\overline{\mathbb{V}}_1+D(1+Y_0 X_0)Q(\p c-\overline{\p}\overline{c})+B(1+Y_0 X_0) Q(\p c+\overline{\p}\overline{c})\nonumber\\
&=  \varepsilon_{ik}(c\mathbb{V}^i_1-e^{\phi}\eta\mathbb{V}^i_\frac{1}{2})\overline{c}\overline{\mathbb{W}}^k_1+\nonumber\\
&\hspace{1cm}-(D+B)(c\p^2 c-3\eta\p^2 \eta e^{2\phi}+2 G c\eta e^{\phi}-8\eta \p \eta\p\phi e^{2\phi}-4 bc\eta \p \eta e^{2\phi} )\nonumber\\
&\hspace{1cm}-(D-B)(\overline{c}\p^2 \overline{c})
% (c + \p c+(1/2)\p^2 c)(c\p c - c\p^2 c+(1/2)c\p^3 c +(1/2)\p c \p^2 c)
\end{align}
\end{subequations}
It is then straightforward to compute that due to the presence of $P_0$, it is actually only $\varphi_A$ that  contributes into the computation $P_0[\eta_0 \varphi_1,Q\varphi_1]$. Let us actually elaborate a bit on this point. While we note that the above expressions for $\eta\varphi_1$ and $Q\varphi_1$ generally contain non-primary contributions, we will see that the ghost and picture conservation pose strong enough constraints so that we will be able to say that everything except for $P_0[\eta_0\varphi_A,Q\varphi_A]$ has to vanish, without actually having to go through explicit computations of the 2-string product. For instance, setting $\eta_0\varphi_1=\eta_0\varphi_A$ and $Q\varphi_1 = -(D+B)c\p^2 c$, we first have
\begin{align}
P_0[\eta_0 \varphi_A,c\p^2 c] \propto  b_0^-\delta(L_0^-)P_0\big(\varepsilon_{ik} c \p c \p^2 c \overline{c} e^{-\phi}(\mathbb{V}_\frac{1}{2})^i(\overline{\mathbb{W}}_1)^k+\ldots \big)\label{eq:WPStart}
\end{align}
where on the r.h.s.\ we have written down explicitly only the term with lowest possible conformal weight which is in principle allowed to appear in the 2-string product. But this state has $(h,\overline{h}) = (1,0)$ so that it is actually killed by both $P_0$ and $\delta(L_0^-)$. We may proceed analogously for the rest of $Q\varphi_1$ while keeping $\eta_0\varphi_1 = \eta_0\varphi_A$: we obtain
\begingroup\allowdisplaybreaks
\begin{subequations}
\begin{align}
P_0[\eta_0 \varphi_A,\eta\p^2 \eta e^{2\phi}] &\propto b_0^-\delta(L_0^-)P_0\big(\varepsilon_{ik} c\overline{c}\eta \p \eta e^{\phi}(\mathbb{V}_\frac{1}{2})^i(\overline{\mathbb{W}}_1)^k+\ldots\big)\,,\\
P_0[\eta_0 \varphi_A,Gc\eta e^{\phi}] &\propto b_0^-\delta(L_0^-)P_0\big(\varepsilon_{ik} c\p c \overline{c}\eta  (\mathbb{V}
_1)^i(\overline{\mathbb{W}}_1)^k+\ldots\big)\,,\\[1mm]
P_0[\eta_0 \varphi_A,\eta \p \eta \p \phi e^{2\phi}] &\propto b_0^-\delta(L_0^-)P_0\big(\varepsilon_{ik} c\overline{c}\eta \p \eta e^{\phi}(\mathbb{V}_\frac{1}{2})^i(\overline{\mathbb{W}}_1)^k+\ldots\big)\,,\\
P_0[\eta_0 \varphi_A,bc\eta \p \eta  e^{2\phi}] &\propto b_0^-\delta(L_0^-)P_0\big(\varepsilon_{ik} c\overline{c}\eta \p \eta e^{\phi}(\mathbb{V}_\frac{1}{2})^i(\overline{\mathbb{W}}_1)^k+\ldots\big)\,,\\
P_0[\eta_0 \varphi_A,\overline{c}\overline{\p}^2\overline{c}] &\propto b_0^-\delta(L_0^-)P_0\big(\varepsilon_{ik} c\overline{c}\overline{\p}\overline{c}\overline{\p}^2\overline{c} e^{-\phi}(\mathbb{V}_\frac{1}{2})^i(\overline{\mathbb{W}}_1)^k+\ldots\big)\,,
\end{align}
\end{subequations}
\endgroup
where none of the expressions inside the parentheses on the r.h.s.\ has conformal weights $(h,\overline{h})=(0,0)$. Hence, all of them are killed by $\delta(L_0^-)P_0$. We may repeat this also for the rest of the terms in $\eta_0 \varphi_1$. Namely, we have
\begingroup\allowdisplaybreaks
\begin{subequations}
\begin{align}
P_0[c\eta,c\p^2 c] &\propto  b_0^-\delta(L_0^-)P_0\big(c\p c\p^2 c \eta+\ldots \big)\,,\\
P_0[c\eta,\eta\p^2 \eta e^{2\phi}] &\propto b_0^-\delta(L_0^-)P_0\big(c\eta \p \eta \p^2 \eta e^{2\phi}+\ldots\big)\,,\\
P_0[c\eta,Gc\eta e^{\phi}] &\propto b_0^-\delta(L_0^-)P_0\big(c\p c \eta \p \eta e^{\phi} G+\ldots\big)\,,\\
P_0[c\eta,\eta \p \eta \p \phi e^{2\phi}] &\propto b_0^-\delta(L_0^-)P_0\big(c\eta \p \eta \p^2 \eta e^{2\phi}+\ldots\big)\,,\\
P_0[c\eta,bc\eta \p \eta  e^{2\phi}] &\propto b_0^-\delta(L_0^-)P_0\big(c\eta \p \eta \p^2 \eta e^{2\phi}+\ldots\big)\,,\\
P_0[c\eta,\overline{c}\overline{\p}^2\overline{c}] &\propto b_0^-\delta(L_0^-)P_0\big(c\eta (\overline{c}\overline{\p}\overline{c}+\alpha  \overline{c}\overline{\p}^2\overline{c})+\ldots\big)\,,
\end{align}
\end{subequations}
\endgroup
where all terms in the parentheses on the r.h.s.\ except for the last line are killed by $\delta(L_0^-)P_0$ and, while the second term of the last line survives $\delta(L_0^-)P_0$ (here $\alpha$ is some number), it is killed by $b_0^-$. Finally, we have
\begingroup\allowdisplaybreaks
\begin{subequations}
\begin{align}
P_0[c \p \xi e^{-2\phi}\overline{c}\overline{\p}^2 \overline{c},c\p^2 c] &\propto  b_0^-\delta(L_0^-)P_0\big(c\p c\p^2 c \p \xi e^{-2\phi}(\overline{c}\overline{\p}\overline{c}+\alpha\overline{c}\overline{\p}^2\overline{c})+\ldots \big)\,,\\
P_0[c \p \xi e^{-2\phi}\overline{c}\overline{\p}^2 \overline{c},\eta\p^2 \eta e^{2\phi}] &\propto b_0^-\delta(L_0^-)P_0\big(c\eta (\overline{c}\overline{\p}\overline{c}+\alpha  \overline{c}\overline{\p}^2\overline{c})+\ldots\big)\,,\\
P_0[c \p \xi e^{-2\phi}\overline{c}\overline{\p}^2 \overline{c},Gc\eta e^{\phi}] &\propto b_0^-\delta(L_0^-)P_0\big(c\p c Ge^{-\phi} (\overline{c}\overline{\p} \overline{c}+\alpha  \overline{c}\overline{\p}^2 \overline{c})+\ldots\big)\,,\\
P_0[c \p \xi e^{-2\phi}\overline{c}\overline{\p}^2 \overline{c},\eta \p \eta \p \phi e^{2\phi}] &\propto b_0^-\delta(L_0^-)P_0\big(c\eta (\overline{c}\overline{\p}\overline{c}+\alpha  \overline{c}\overline{\p}^2\overline{c})+\ldots\big)\,,\\
P_0[c \p \xi e^{-2\phi}\overline{c}\overline{\p}^2 \overline{c},bc\eta \p \eta  e^{2\phi}] &\propto b_0^-\delta(L_0^-)P_0\big(c\eta (\overline{c}\overline{\p}\overline{c}+\alpha  \overline{c}\overline{\p}^2\overline{c})+\ldots\big)\,,\\
P_0[c \p \xi e^{-2\phi}\overline{c}\overline{\p}^2 \overline{c},\overline{c}\overline{\p}^2\overline{c}] &\propto b_0^-\delta(L_0^-)P_0\big(c\p \xi e^{-2\phi} \overline{c}\overline{\p}\overline{c}\overline{\p}^2 \overline{c}\overline{\p}^3 \overline{c}+\ldots\big)\,,
\end{align}
\end{subequations}
\endgroup
where, again, all terms on the r.h.s. are killed by $b_0^-\delta(L_0^-)P_0$.
We can therefore write
\begin{subequations}
\label{eq:weakProj}
\begin{align}
P_0[\eta_0 \varphi_1,Q\varphi_1]
&=P_0[\eta_0 \varphi_A,Q\varphi_A] \\[+1mm]
&=  \varepsilon_{ik} \varepsilon_{jl}b_0^- (-c\p c)\{\mathbb{V}^i_\frac{1}{2}\mathbb{V}_1^j\}_1 e^{-\phi}\overline{c}\overline{\p}\overline{c} \{\overline{\mathbb{V}}^k_1\overline{\mathbb{V}}_1^l\}_1\\
&= - \varepsilon_{ik} \varepsilon_{jl}2b_0^- \p c^+ \p c^- c\{\mathbb{V}^i_\frac{1}{2}\mathbb{V}^j_1\}_1 e^{-\phi}\overline{c} \{\overline{\mathbb{V}}^k_1\overline{\mathbb{V}}^l_1\}_1\\
&= +2 \varepsilon_{ik} \varepsilon_{jl}  c_0^+  c\{\mathbb{V}_\frac{1}{2}^i\mathbb{V}_1^j\}_1 e^{-\phi}\overline{c} \{\overline{\mathbb{V}}^k_1\overline{\mathbb{V}}^l_1\}_1
\end{align}
\end{subequations}
and therefore
\begin{align}
c_0^+ P_0[\eta_0 \varphi_1,Q\varphi_1]=0\,.
\end{align}
This shows that we actually have
\begin{align}
\varphi_2(\varphi_1) = \mathcal{O}(\varphi_1^3)\,.
\end{align}
Because the cross term $\big\langle \eta_0 \varphi_1,Q\varphi_2\big\rangle$ in the quadratic part of the effective action vanishes, this means that integrating out $\varphi_2$ may only yield corrections to $\mathcal{O}(\varphi_1^5)$ couplings. We may therefore write the effective action for $\varphi_1$ as
\begin{align}
S_{\text{eff},1}(\varphi_1) &= \frac{1}{3!}\big\langle\eta_0 \varphi_1,[\varphi_1,Q\varphi_1]\big\rangle+\nonumber\\
&\hspace{1.5cm}+\frac{1}{4!}\bigg(\big\langle \eta_0 \varphi_1,[\varphi_1,Q\varphi_1,Q\varphi_1]\big\rangle+\big\langle\eta_0 \varphi_1,[\varphi_1,[\varphi_1,Q\varphi_1]]\big\rangle\bigg)+\nonumber\\
&\hspace{4cm}
+\frac{1}{8}\bigg\langle [\eta_0 \varphi_1,Q\varphi_1],\frac{b_0^+}{L_0^+}\xi_0 \overline{P}_0[\eta_0 \varphi_1,Q\varphi_1]\bigg\rangle+\mathcal{O}(\varphi_1^5)\,.
\end{align}
This may be succinctly rewritten in the small Hilbert space: substituting $\varphi_1=\xi_0 \psi_1$ where $\eta_0\psi_1=0$, we have $\eta_0\varphi_1= \psi_1$ together with $Q\varphi_1=X_0 \psi_1$ so that
\begin{align}
S_{\text{eff},S,1}(\psi_1) &= \frac{1}{3!}\big\langle\psi_1,[\psi_1,X_0\psi_1]\big\rangle_S+\nonumber\\
&\hspace{1cm}+\frac{1}{4!}\bigg(\big\langle \psi_1,[ \psi_1,X_0\psi_1,X_0 \psi_1]\big\rangle_S+\nonumber\\
&\hspace{2cm}+\big\langle\psi_1,[\xi_0\psi_1,[\psi_1,X_0 \psi_1]+[\psi_1,[\xi_0\psi_1,X_0 \psi_1]]\big\rangle_S\bigg)+\nonumber\\
&\hspace{3cm}
-\frac{1}{8}\bigg\langle [\psi_1,X_0 \psi_1],\frac{b_0^+}{L_0^+} \overline{P}_0[\psi_1,X_0 \psi_1]\bigg\rangle_S+\mathcal{O}(\psi_1^5)\,.
\end{align}
\subsection{Decoupling of pure gauge physical state}
Finally, at this point we also note that we can remove the trivial element $\varphi_B$ of the semi-relative cohomology. Indeed, we note that we may write
\begin{align}
\varphi_B = \xi_0 Q \chi_B\,,
\end{align}
or, equivalently $\varphi_B = \xi_0 \psi_B$ where $\psi_B = Q\chi_B$ with $b_0^- \chi_B= L_0^-\chi_B=\eta_0 \chi_B=0$. We can then decompose
\begin{align}
\varphi_1 = \varphi_\text{p}+\varphi_B
\end{align}
where we have denoted $\varphi_\text{p}=\varphi_A+\varphi_D$ the physical excitations (equivalently, we have $\psi_1 = \psi_\text{p}+\psi_B$ where $\eta_0\psi_\text{p}=\eta_0\psi_B=0$ with $\varphi_\text{p}=\xi_0\psi_\text{p}$ and $\varphi_B=\xi_0\psi_B$ so that $\psi_B = Q\chi_B$). Let us also write
\begin{align}
%S_{\text{eff},1}(\varphi_1)&=\sum_{k=3}^\infty S_{\text{eff},1}^{(k)}(\varphi_1)\,,\\
S_{\text{eff},S,1}(\psi_1)&=\sum_{k=3}^\infty S_{\text{eff},S,1}^{(k)}(\psi_1)
\end{align}
where $S_{\text{eff},S,1}^{(k)}$ contains $k$ powers of $\psi_1$. Using that $Q\psi_1=Q\psi_\text{p}=0$, we have (decomposing the three insertions as $\psi_1 = \psi_\text{p}+Q\chi_B$ one by one)
\begingroup\allowdisplaybreaks
\begin{subequations}
\begin{align}
S_{\text{eff},1,S}^{(3)}(\psi_1) &= \frac{1}{3!}\big\langle \psi_\text{p}+Q\chi_B,[\psi_1,X_0\psi_1]\big\rangle\\
&= \frac{1}{3!}\big\langle \psi_\text{p},[\psi_1,X_0\psi_1]\big\rangle+\nonumber\\
&\hspace{2cm}-\frac{1}{3!}\big\langle \chi_B,Q[\psi_1,X_0\psi_1]\big\rangle\\
&= \frac{1}{3!}\big\langle \psi_\text{p},[\psi_1,X_0\psi_1]
%&= \frac{1}{3!}\big\langle \psi_\text{p},[\psi_\text{p},X_0\psi_1]\big\rangle+\nonumber\\
%&\hspace{2cm}+ \frac{1}{3!}\big\langle \psi_\text{p},[Q\chi_B,X_0\psi_1]\big\rangle\\
%&= \frac{1}{3!}\big\langle \psi_\text{p},[\psi_\text{p},X_0\psi_\text{p}]\big\rangle+\frac{1}{3!}\big\langle \psi_\text{p},[\psi_\text{p},X_0Q\chi_B]\big\rangle\\
%&= \frac{1}{3!}\big\langle \psi_\text{p},[\psi_\text{p},X_0\psi_\text{p}]\big\rangle\\[+1mm]
%&=S_{\text{eff},1,S}^{(3)}(\psi_\text{p})\,.
\end{align}
\end{subequations}
\endgroup
and similarly for the remaining two insertions. At the end we find that
\begin{align}
S_{\text{eff},1,S}^{(3)}(\psi_1)=S_{\text{eff},1,S}^{(3)}(\psi_\text{p})\,.
\end{align}
For the quartic coupling, we have (again, focusing on decomposing the first insertion)
\begingroup\allowdisplaybreaks
\begin{subequations}
\begin{align}
S_{\text{eff},1,S}^{(4)}(\psi_1)&=\frac{1}{4!}\bigg(\big\langle \psi_\text{p}+\psi_B,[ \psi_1,X_0\psi_1,X_0 \psi_1]\big\rangle_S+\nonumber\\
&\hspace{1cm}+\big\langle\psi_\text{p}+\psi_B,[\xi_0\psi_1,[\psi_1,X_0 \psi_1]+[\psi_\text{p}+\psi_B,[\xi_0\psi_1,X_0 \psi_1]]\big\rangle_S\bigg)+\nonumber\\
&\hspace{3cm}
-\frac{1}{8}\bigg\langle [\psi_\text{p}+\psi_B,X_0 \psi_1],\frac{b_0^+}{L_0^+} \overline{P}_0[\psi_1,X_0 \psi_1]\bigg\rangle_S\\
%&=
%\frac{1}{4!}\bigg(\big\langle \psi_\text{p},[ \psi_1,X_0\psi_1,X_0 \psi_1]\big\rangle_S+\nonumber\\
%&\hspace{1cm}+\big\langle\psi_\text{p},[\xi_0\psi_1,[\psi_1,X_0 \psi_1]+[\psi_\text{p},[\xi_0\psi_1,X_0 \psi_1]]\big\rangle_S\bigg)+\nonumber\\
%&\hspace{3cm}
%-\frac{1}{8}\bigg\langle [\psi_\text{p},X_0 \psi_1],\frac{b_0^+}{L_0^+} \overline{P}_0[\psi_1,X_0 \psi_1]\bigg\rangle_S\\
%&\hspace{0.4cm}
%+\frac{1}{4!}\bigg(\big\langle Q\chi_B,[ \psi_1,X_0\psi_1,X_0 \psi_1]\big\rangle_S+\nonumber\\
%&\hspace{1cm}+\big\langle Q\chi_B,[\xi_0\psi_1,[\psi_1,X_0 \psi_1]+[\psi_1,[\xi_0\psi_1,X_0 \psi_1]]\big\rangle_S\bigg)+\nonumber\\
%&\hspace{3cm}
%-\frac{1}{8}\bigg\langle [Q\chi_B,X_0 \psi_1],\frac{b_0^+}{L_0^+} \overline{P}_0[\psi_1,X_0 \psi_1]\bigg\rangle_S\\
&=
\frac{1}{4!}\bigg(\big\langle \psi_\text{p},[ \psi_1,X_0\psi_1,X_0 \psi_1]\big\rangle_S+\nonumber\\
&\hspace{1cm}+\big\langle\psi_\text{p},[\xi_0\psi_1,[\psi_1,X_0 \psi_1]+[\psi_\text{p},[\xi_0\psi_1,X_0 \psi_1]]\big\rangle_S\bigg)+\nonumber\\
&\hspace{3cm}
-\frac{1}{8}\bigg\langle [\psi_\text{p},X_0 \psi_1],\frac{b_0^+}{L_0^+} \overline{P}_0[\psi_1,X_0 \psi_1]\bigg\rangle_S\\
&\hspace{0.4cm}
+\frac{1}{4!}\bigg(
2\big\langle \chi_B,[[ \psi_1,X_0\psi_1],X_0 \psi_1]\big\rangle_S
+\big\langle \chi_B,[[X_0 \psi_1,X_0\psi_1], \psi_1]\big\rangle_S
+\nonumber\\
&\hspace{1cm}+\big\langle \chi_B,[X_0\psi_1,[\psi_1,X_0 \psi_1]\big\rangle_S-\big\langle \chi_B, [\psi_1,[X_0\psi_1,X_0 \psi_1]]\big\rangle_S\bigg)+\nonumber\\
&\hspace{3cm}
-\frac{1}{8}\big\langle \chi_B,X_0 [\psi_1, \overline{P}_0[\psi_1,X_0 \psi_1]]\big\rangle_S\\
&=
\frac{1}{4!}\bigg(\big\langle \psi_\text{p},[ \psi_1,X_0\psi_1,X_0 \psi_1]\big\rangle_S+\nonumber\\
&\hspace{1cm}+\big\langle\psi_\text{p},[\xi_0\psi_1,[\psi_1,X_0 \psi_1]+[\psi_\text{p},[\xi_0\psi_1,X_0 \psi_1]]\big\rangle_S\bigg)+\nonumber\\
&\hspace{3cm}
-\frac{1}{8}\bigg\langle [\psi_\text{p},X_0 \psi_1],\frac{b_0^+}{L_0^+} \overline{P}_0[\psi_1,X_0 \psi_1]\bigg\rangle_S\\
&\hspace{0.4cm}
+\frac{1}{4!}\bigg(
3\big\langle \chi_B,[[ \psi_1,X_0\psi_1],X_0 \psi_1]\big\rangle_S
+\nonumber\\
&\hspace{3cm}-3\big\langle \chi_B,X_0 [\psi_1, [\psi_1,X_0 \psi_1]]\big\rangle_S\bigg)+\nonumber\\
&\hspace{5cm}
+\frac{1}{8}\big\langle \chi_B,X_0 [\psi_1, P_0[\psi_1,X_0 \psi_1]]\big\rangle_S\\
&=
\frac{1}{4!}\bigg(\big\langle \psi_\text{p},[ \psi_1,X_0\psi_1,X_0 \psi_1]\big\rangle_S+\nonumber\\
&\hspace{1cm}+\big\langle\psi_\text{p},[\xi_0\psi_1,[\psi_1,X_0 \psi_1]+[\psi_\text{p},[\xi_0\psi_1,X_0 \psi_1]]\big\rangle_S\bigg)+\nonumber\\
&\hspace{3cm}
-\frac{1}{8}\bigg\langle [\psi_\text{p},X_0 \psi_1],\frac{b_0^+}{L_0^+} \overline{P}_0[\psi_1,X_0 \psi_1]\bigg\rangle_S,
\end{align}
\end{subequations}
\endgroup
where in the second step we have used one of the $L_\infty$ relations to move $Q$ inside the 3-string product and in the last step, we have recalled that
\begin{align}
P_0[\chi_B,X_0\psi_1] = P_0[\eta \varphi_B, Q\varphi_1]=0\label{eq:weakProjChiPsi}
\end{align}
which we have noticed when evaluating \eqref{eq:weakProj}. It is straightforward to continue in the same fashion to show that we actually have
\begin{align}
S_{\text{eff},1,S}^{(4)}(\psi_1)&=S_{\text{eff},1,S}^{(4)}(\psi_\text{p})\,.
\end{align}
%It would be very interesting to investigate in more detail whether the ghost dilaton completely, or at least partially, drops out also from the quartic part of the effective action. We hope to come back to investigating such a ``Dilaton Theorem'' for the heterotic string in a later work (see CITE:YangANDZwiebach for the bosonic case).
%
%%%%%%%%%%%%%%%%%%%%%%
\section{Vanishing of the cubic potential for marginal (anti)chiral-ring states}\label{app:C}
%is %%%%%%%%%%%%%%%%%%%%
In this appendix we give a proof of \eqref{eq:projTest}, when the involved marginal matter field can be decomposed in $q=\pm 1$ charged $\N=2$ states. In other words, we are going to assume that the marginal matter field consists purely of states from the chiral and anti-chiral ring of the $\mathcal{N}=2$ SCFT. First, let us recall the generalized Wick theorem
\begin{align}
\contraction{}{A(z)}{(B }{}
\contraction{A(z)(B C)(w)=\frac{1}{2\pi i}\oint_w \frac{dx}{x-w}\bigg\{}{A(z)}{}{B(x)}
\contraction{A(z)(B C)(w)=\frac{1}{2\pi i}\oint_w \frac{dx}{x-w}\bigg\{
A(z)B(x)C(w)+B(x)}{A(z)}{}{C(w)}
A(z)(B C)(w)=\frac{1}{2\pi i}\oint_w \frac{dx}{x-w}\bigg\{
A(z)B(x)C(w)+B(x)A(z)C(w)
\bigg\}
\end{align}
where $\contraction{}{A(z)}{}{B(w)}A(z)B(w)$ denotes the contraction of local fields $A(z)$ and $B(w)$, and $(AB)$ their normal-ordered product. That is
\begin{align}
\contraction{A(z)B(w)=}{A(z)}{}{B(w)}
A(z)B(w)=A(z)B(w)+(AB)(w)+\mathcal{O}[(z-w)^1]\,.
\end{align}
We can apply this result to first show that
\begingroup\allowdisplaybreaks
\begin{subequations}
\begin{align}
\contraction{}{G^\pm(z)}{\big((\mathbb{V}_\frac{1}{2}^i)^\pm }{}
\contraction{G^\pm(z)\big((\mathbb{V}_\frac{1}{2}^i)^\pm (\mathbb{V}_\frac{1}{2}^j)^\mp\big)(w)=\frac{1}{2\pi i}\oint_w \frac{dx}{x-w}\bigg\{}{G^\pm(z)}{}{(\mathbb{V}_\frac{1}{2}^i)^\pm(x)}
G^\pm(z)\big((\mathbb{V}_\frac{1}{2}^i)^\pm (\mathbb{V}_\frac{1}{2}^j)^\mp\big)(w)&=\frac{1}{2\pi i}\oint_w \frac{dx}{x-w}\bigg\{
G^\pm(z)(\mathbb{V}_\frac{1}{2}^i)^\pm(x)(\mathbb{V}_\frac{1}{2}^j)^\mp(w)+\nonumber\\
\contraction{\hspace{4cm}+(\mathbb{V}_\frac{1}{2}^i)^\pm(x)}{G^\pm(z)}{}{(\mathbb{V}_\frac{1}{2}^j)^\mp(w)}
&\hspace{4cm}+(\mathbb{V}_\frac{1}{2}^i)^\pm(x)G^\pm(z)(\mathbb{V}_\frac{1}{2}^j)^\mp(w)
\bigg\}\\
&=
\contraction{\frac{1}{2\pi i}\oint \frac{dx}{x-w} (\mathbb{V}_\frac{1}{2}^i)^\pm(x)}{G^\pm(z)}{}{(\mathbb{V}_\frac{1}{2}^j)^\mp(w)}
\frac{1}{2\pi i}\oint_w \frac{dx}{x-w} (\mathbb{V}_\frac{1}{2}^i)^\pm(x)
G^\pm(z)(\mathbb{V}_\frac{1}{2}^j)^\mp(w)\\
&=
\frac{1}{2\pi i}\oint_w \frac{dx}{(x-w)(z-w)} (\mathbb{V}_\frac{1}{2}^i)^\pm(x)
(\mathbb{V}_1^j)^\mp(w)\\
&=
(z-w)^{-1}\big\{(\mathbb{V}_\frac{1}{2}^i)^\pm
(\mathbb{V}_1^j)^\mp\big\}_0(w)\,.
\end{align}
\end{subequations}
\endgroup
The upshot is that we have
\begin{subequations}
\begin{align}
G^\pm_{\frac{1}{2}}\big\{(\mathbb{V}_\frac{1}{2}^i)^\pm(\mathbb{V}_\frac{1}{2}^j)^\mp\big\}_0&=0\,.
\end{align}
\end{subequations}
This means that for any $(\mathbb{V}_{\frac{1}{2}}^i)^\pm$,$(\mathbb{V}_{\frac{1}{2}}^j)^\pm$, $(\mathbb{V}_{\frac{1}{2}}^k)^\pm$, we have
%\begin{subequations}
\begin{align}
%\big\langle (\mathbb{V}_{1}^i)^\mp(z_1)(\mathbb{V}_{\frac{1}{2}}^j)^\pm(z_2)(\mathbb{V}_{\frac{1}{2}}^k)^\mp(z_3)\big\rangle
%&= \frac{\big\langle G_{-\frac{1}{2}}^\pm (\mathbb{V}_{\frac{1}{2}}^i)^\mp \big| \big\{(\mathbb{V}_{\frac{1}{2}}^j)^\pm(\mathbb{V}_{\frac{1}{2}}^k)^\mp\big\}_0\big\rangle}{(z_1-z_2)(z_1-z_3)}\\
%&= \frac{\big\langle  (\mathbb{V}_{\frac{1}{2}}^i)^\mp \big|G_{\frac{1}{2}}^\pm \big\{(\mathbb{V}_{\frac{1}{2}}^j)^\pm(\mathbb{V}_{\frac{1}{2}}^k)^\mp\big\}_0\big\rangle}{(z_1-z_2)(z_1-z_3)}
\big\langle G_{-\frac{1}{2}}^\pm (\mathbb{V}_{\frac{1}{2}}^i)^\mp \big| \big\{(\mathbb{V}_{\frac{1}{2}}^j)^\pm(\mathbb{V}_{\frac{1}{2}}^k)^\mp\big\}_0\big\rangle
&= {\big\langle  (\mathbb{V}_{\frac{1}{2}}^i)^\mp \big|G_{\frac{1}{2}}^\pm \big\{(\mathbb{V}_{\frac{1}{2}}^j)^\pm(\mathbb{V}_{\frac{1}{2}}^k)^\mp\big\}_0\big\rangle}=0\,.\label{eq:CorPMPM}
\end{align}
%\end{subequations}
We also have
\begingroup\allowdisplaybreaks
\begin{subequations}
\label{eq:con2}
\begin{align}
\contraction{}{G^\pm(z)}{\big((\mathbb{V}_\frac{1}{2}^i)^\mp }{}
\contraction{G^\pm(z)\big((\mathbb{V}_\frac{1}{2}^i)^\mp (\mathbb{V}_\frac{1}{2}^j)^\pm\big)(w)=\frac{1}{2\pi i}\oint_w \frac{dx}{x-w}\bigg\{}{G^\pm(z)}{}{(\mathbb{V}_\frac{1}{2}^i)^\mp(x)}
G^\pm(z)\big((\mathbb{V}_\frac{1}{2}^i)^\mp (\mathbb{V}_\frac{1}{2}^j)^\pm\big)(w)&=\frac{1}{2\pi i}\oint_w \frac{dx}{x-w}\bigg\{
G^\pm(z)(\mathbb{V}_\frac{1}{2}^i)^\mp(x)(\mathbb{V}_\frac{1}{2}^j)^\pm(w)+\nonumber\\
\contraction{\hspace{4cm}+(\mathbb{V}_\frac{1}{2}^i)^\mp(x)}{G^\pm(z)}{}{(\mathbb{V}_\frac{1}{2}^j)^\pm(w)}
&\hspace{4cm}+(\mathbb{V}_\frac{1}{2}^i)^\mp(x)G^\pm(z)(\mathbb{V}_\frac{1}{2}^j)^\pm(w)
\bigg\}\\
&=
\contraction{\frac{1}{2\pi i}\oint \frac{dx}{x-w}}{G^\pm(z)}{}{(\mathbb{V}_\frac{1}{2}^i)^\mp(x)}
\frac{1}{2\pi i}\oint_w \frac{dx}{x-w}
G^\pm(z)(\mathbb{V}_\frac{1}{2}^i)^\mp(x)(\mathbb{V}_\frac{1}{2}^j)^\pm(w)\\
&=
\frac{1}{2\pi i}\oint_w \frac{dx}{(x-w)(z-x)}
(\mathbb{V}_1^i)^\mp(x)(\mathbb{V}_\frac{1}{2}^j)^\pm(w)\\[+1.5mm]
&=
(z-w)^{-1}\big\{(\mathbb{V}_1^i)^\mp
(\mathbb{V}_\frac{1}{2}^j)^\pm\big\}_0\,,
\end{align}
\end{subequations}
\endgroup
where in the last step, we have used \eqref{eq:CorPMPM}.
Again, the upshot is that we have
\begin{subequations}
\begin{align}
G^\pm_{\frac{1}{2}}\big\{(\mathbb{V}_\frac{1}{2}^i)^\mp(\mathbb{V}_\frac{1}{2}^j)^\pm\big\}_0&=0\,.
\end{align}
\end{subequations}
This means that for any $(\mathbb{V}_{\frac{1}{2}}^i)^\pm$,$(\mathbb{V}_{\frac{1}{2}}^j)^\pm$, $(\mathbb{V}_{\frac{1}{2}}^k)^\pm$, we have
%\begin{subequations}
\begin{align}
\big\langle G_{-\frac{1}{2}}^\pm (\mathbb{V}_{\frac{1}{2}}^i)^\mp \big| \big\{(\mathbb{V}_{\frac{1}{2}}^j)^\mp(\mathbb{V}_{\frac{1}{2}}^k)^\pm\big\}_0\big\rangle
&= {\big\langle  (\mathbb{V}_{\frac{1}{2}}^i)^\mp \big|G_{\frac{1}{2}}^\pm \big\{(\mathbb{V}_{\frac{1}{2}}^j)^\mp(\mathbb{V}_{\frac{1}{2}}^k)^\pm\big\}_0\big\rangle}=0\,.
\end{align}
Altogether, we therefore obtain
\begin{subequations}
\begin{align}
\big\langle \mathbb{V}^i_\frac{1}{2}\big|\big\{\mathbb{V}^{j}_\frac{1}{2}(G_{-\frac{1}{2}}\mathbb{V}_\frac{1}{2}^k)\big\}_1\big\rangle&=\big\langle (G_{-\frac{1}{2}}\mathbb{V}_\frac{1}{2}^k)\big|\big\{\mathbb{V}^i_\frac{1}{2}\mathbb{V}^{j}_\frac{1}{2}\big\}_0\big\rangle\\
&=\big\langle G_{-\frac{1}{2}}^+(\mathbb{V}_\frac{1}{2}^k)^-\big|\big\{(\mathbb{V}^i_\frac{1}{2})^+(\mathbb{V}^{j}_\frac{1}{2})^-\big\}_0\big\rangle
+\nonumber\\
&\hspace{2cm}+\big\langle G_{-\frac{1}{2}}^-(\mathbb{V}_\frac{1}{2}^k)^+\big|\big\{(\mathbb{V}^i_\frac{1}{2})^-(\mathbb{V}^{j}_\frac{1}{2})^+\big\}_0\big\rangle\nonumber\\
&\hspace{1cm}+\big\langle G_{-\frac{1}{2}}^+(\mathbb{V}_\frac{1}{2}^k)^-\big|\big\{(\mathbb{V}^i_\frac{1}{2})^-(\mathbb{V}^{j}_\frac{1}{2})^+\big\}_0\big\rangle
+\nonumber\\
&\hspace{2cm}+\big\langle G_{-\frac{1}{2}}^-(\mathbb{V}_\frac{1}{2}^k)^+\big|\big\{(\mathbb{V}^i_\frac{1}{2})^+(\mathbb{V}^{j}_\frac{1}{2})^-\big\}_0\big\rangle\\
&=0\,.
\end{align}
\end{subequations}
This shows that the cubic coupling in the effective action vanishes whenever we have the $\mathcal{N}=2$ decomposition. Note that since there are no Chan-Paton factors in the heterotic string theory, we could have noted that
\begin{align}
\big\{(\mathbb{V}^i_\frac{1}{2})^+(\mathbb{V}^{j}_\frac{1}{2})^-\big\}_0 = -\big\{(\mathbb{V}^{j}_\frac{1}{2})^-(\mathbb{V}^i_\frac{1}{2})^+\big\}_0
\end{align}
and therefore did not have to go through explicitly computing the second contraction \eqref{eq:con2}. However, doing so we have exhibited that our derivation would have worked even in the open superstring case, as it was claimed in \cite{Vosmera:2019mzw}. Also note that we have in fact shown a somewhat stronger result because \eqref{eq:projTest}, being valid for all $\mathbb{V}^{i}_\frac{1}{2}$ implies that
\begin{align}
\big\{\mathbb{V}^{j}_\frac{1}{2}(G_{-\frac{1}{2}}\mathbb{V}_\frac{1}{2}^k)\big\}_1=0,
\end{align}
whenever we have $\mathcal{N}=2$ decomposition.

\end{document}